\documentclass[12pt,titlepage]{article}
\usepackage[letterpaper,top=1in,bottom=1in,left=1.25in,right=1.25in]{geometry}
\usepackage{graphicx,cite, url}
\usepackage{color}
\usepackage{threeparttable}
\usepackage{amsmath,amssymb}
\usepackage{makecell}
\usepackage{multirow}
\usepackage{float}
\usepackage{hyperref}

\begin{document}

\title{Experimental robustness of Fourier Ptychography phase retrieval algorithms}

\author{Li-Hao Yeh,$^1$ Jonathan Dong,$^{1,2}$ Jingshan Zhong,$^1$ Lei Tian,$^1$\\ Michael Chen,$^1$ Gongguo Tang,$^3$ Mahdi Soltanolkotabi,$^4$ and Laura Waller,$^{1,*}$\\
\\
\multicolumn{1}{p{\textwidth}}{\centering\emph{\normalsize 1. Department of Electrical Engineering and Computer Sciences, University of California, Berkeley, 94720, USA\\
2. D\'{e}partement de Physique, \'{E}cole Normale Sup\'{e}rieure, Paris 75005, France\\
3. Department of Electrical Engineering and Computer Sciences, Colorado School of Mines, Golden, 80401, USA\\
4. Department of Electrical Engineering, University of Southern California, Los Angeles, 90089, USA\\
$^{*}$  waller@berkeley.edu
}}}

\maketitle

\begin{abstract}
Fourier ptychography is a new computational microscopy technique that provides gigapixel-scale intensity and phase images with both wide field-of-view and high resolution. By capturing a stack of low-resolution images under different illumination angles, an inverse algorithm can be used to computationally reconstruct the high-resolution complex field. Here, we compare and classify multiple proposed inverse algorithms in terms of experimental robustness. We find that the main sources of error are noise, aberrations and mis-calibration (i.e. \textit{model mis-match}). Using simulations and experiments, we demonstrate that the choice of cost function plays a critical role, with \textit{amplitude-based} cost functions performing better than \textit{intensity-based} ones. The reason for this is that Fourier ptychography datasets consist of images from both brightfield and darkfield illumination, representing a large range of measured intensities. Both noise (e.g. Poisson noise) and model mis-match errors are shown to scale with intensity. Hence, algorithms that use an appropriate cost function will be more tolerant to both noise and model mis-match. Given these insights, we propose a global Newton's method algorithm which is robust and accurate. Finally, we discuss the impact of procedures for algorithmic correction of aberrations and mis-calibration. 
\end{abstract}

\section{Introduction}

Fourier ptychographic microscopy (FPM)~\cite{Zheng:2013gq} circumvents optical space-bandwidth (SBP) limitations to achieve gigapixel-scale quantitative phase images, having both wide field-of-view (FOV) and high resolution. The method combines ideas from synthetic aperture and translational-diversity phase retrieval~\cite{rodenburg2004phase,Brady2009}, conveniently realized by replacing the light source of a microscope with an LED array, then capturing multiple images under different illumination angles. When LEDs illuminate the sample from angles smaller than that allowed by the objective's numerical aperture ($NA_{\mathrm{obj}}$), brightfield images result. Conversely, when the illumination NA is larger than the objective NA, darkfield images result. Although darkfield images alone do not have higher resolution than the objective allows, they do contain information about sub-diffraction-limit sized features, which occupy a shifted area of the sample's Fourier space (assuming a thin sample). By collecting many images that cover a wide region of Fourier space and stitching them together coherently, one can achieve spatial resolution beyond the objective's diffraction limit, corresponding to the sum of illumination and objective NAs ($NA_{\mathrm{eff}} = NA_{\mathrm{illu}} + NA_{\mathrm{obj}}$). FPM's scan-free high SBP imaging capability has great potential for revolutionizing biomedical imaging, with applications in digital pathology~\cite{Zheng:2013gq, Williams2014, Horstmeyer2015, Chung2015} and \emph{in vivo} live cell imaging~\cite{Tian2015a}. The original FPM method only applies to 2D thin objects, however, new models and reconstruction algorithms also enable 3D reconstruction of thick samples~\cite{Tian2015}. 

Multiple algorithms have been proposed for solving the nonlinear inverse FPM problem, which amounts to phase retrieval. Amongst these, there are the usual trade-offs between accuracy, noise performance and computational complexity. In practice, however, we have found that a critical metric is an algorithm's performance under model mis-match -- when the experimental data is imperfect (e.g. due to misalignment). This is typical in computational imaging algorithms, which are often fragile and not robust enough to provide consistent high-quality results. Unfortunately, model mis-match is difficult to quantify, since it is systematic, yet unpredictable. Here, we aim to compare algorithms directly, in order to identify error mechanisms and determine the most accurate and robust algorithm for our experiments. 

\begin{figure} [H]
\centering
\vspace{-5mm}
\includegraphics[width=14cm]{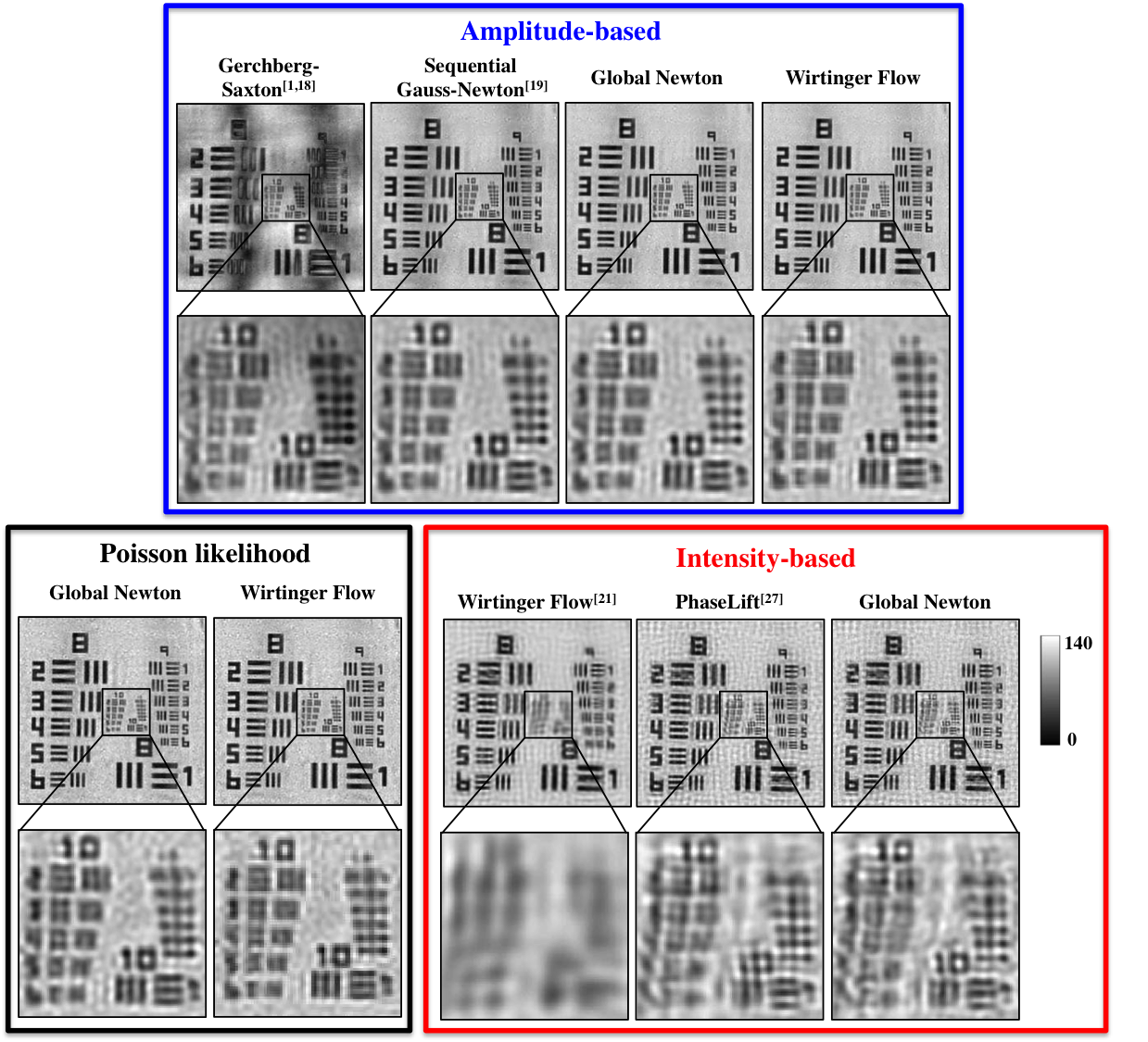}
\vspace{-5mm}
\caption{Fourier ptychographic reconstruction (amplitude only) of a test object with the algorithms discussed here, all using the same experimental dataset. Algorithms derived from the same cost function (amplitude-based, intensity-based, and Poisson-likelihood) give similar performance, and first-order methods (Gerchberg-Saxton) suffer artifacts.}
\label{Fig4}
\end{figure}

The original FPM algorithm used a Gerchberg-Saxton approach~\cite{Gerchberg1971phase}, which is a type of alternating projections~\cite{Fienup1978, Fienup1982, elser03, Marchesini2007}, first developed for traditional ptychography~\cite{rodenburg2004phase, guizar2008phase, Maiden:2009bx, thibault2009probe, Yang:2011ty, Brady2009} and later for FPM~\cite{Zheng:2013gq,Ou:14,Tian2014}. Shifted support constraints (finite pupil size) are enforced in the Fourier domain as the corresponding amplitude constraints (measured images) are applied in the image domain, while letting the phase evolve as each image is stepped through sequentially. The Gerchberg-Saxton method, which is a type of gradient descent, represents a natural way to solve phase retrieval problems by trying to directly minimize some cost function that describes the differences between actual and predicted measurements. Unfortunately, these formulations are often non-convex in nature and do not come with global convergence guarantees. 

Recently, a class of gradient descent like updates, dubbed Wirtinger Flows~\cite{Candes:2014ur}, have been shown to have global convergence guarantees. This method has been successfully applied to FPM~\cite{bian2015fourier}, though the actual implementation deviates from theory somewhat. In the Wirtinger Flow framework, the optimization procedure is similar to gradient descent, except that the step size and initial guess are carefully chosen for provable convergence.

Gradient descent and Wirtinger Flow are \emph{first-order} methods, in the sense that they only use the first-order derivative of the cost function when updating the complex-field. It is also possible, and generally advantageous, to use higher-order derivatives in the updates. For example, \textit{second-order} methods (e.g. Newton's method) use both the first and second derivative of the cost function, and have been shown to provide faster convergence rates~\cite{Nocedal:2006uv}. In our studies, we also observe improved performance when using second-order methods. For example, in the top row of Fig.~\ref{Fig4}, the Gerchberg-Saxton algorithm is a first-order method, whereas the other three methods are second-order (or approximate second-order) methods. All results achieve a similar resolution, but the first-order (Gerchberg-Saxton) result is corrupted by low-frequency artifacts. While computing second-order derivatives increases complexity, we find that it usually reduces the number of iterations needed, enabling fast overall run times.

The final class of algorithms that have been proposed are based on convex relaxations~\cite{Recht:2010ht,Candes:2013ka,candes2013phaselift,Burer:2003fg,horstmeyer2014solving}. This class of phase retrieval algorithms, called PhaseLift, re-frames the problem in higher dimensions such that it becomes convex, then aims to minimize the cost function between actual and predicted intensity via semidefinite programming. These algorithms come with the significant advantage of rigorous mathematical guarantees~\cite{Candes:2013vu} and were successfully applied to FPM data~\cite{horstmeyer2014solving}. The actual implementations of these algorithms, however, deviate from the provable case due to computational limitations. 

Algorithms can be further classified as sequential or global, depending on whether the update is done for each image, one at a time (sequentially), or all at once with the full set of images (globally) for each iteration. Global methods are expected to perform better, at a cost of additional computational requirements. In our studies, results show little difference between the sequential and global implementation of any particular algorithm (see Fig.~\ref{Fig4}), suggesting that sequential procedures may be sufficient, allowing reduced computational requirements. 

One seemingly unimportant classification of algorithms is whether their cost function minimizes differences in \textit{intensity} or \textit{amplitude}. Throughout this paper, we refer to algorithms that minimize intensity differences as \textit{intensity-based} algorithms, and algorithms that minimize amplitude differences as \textit{amplitude-based} algorithms. Since intensity is amplitude squared, both drive the optimization in the correct direction; hence, one might expect that the choice between the two is of little consequence. Surprisingly, however, we find that the cost function is the key predictor of experimental performance for our experimental dataset. \textit{Intensity-based} algorithms suffer from strong artifacts (see Fig.~\ref{Fig4}), which we show to be due to noise and model mis-match errors. Hence, amplitude-based algorithms perform better on imperfect data, so are more robust. Our goal is to explain why this happens.

\begin{figure}[H]
\centering
\vspace{-5mm}
\includegraphics[width=13.1cm]{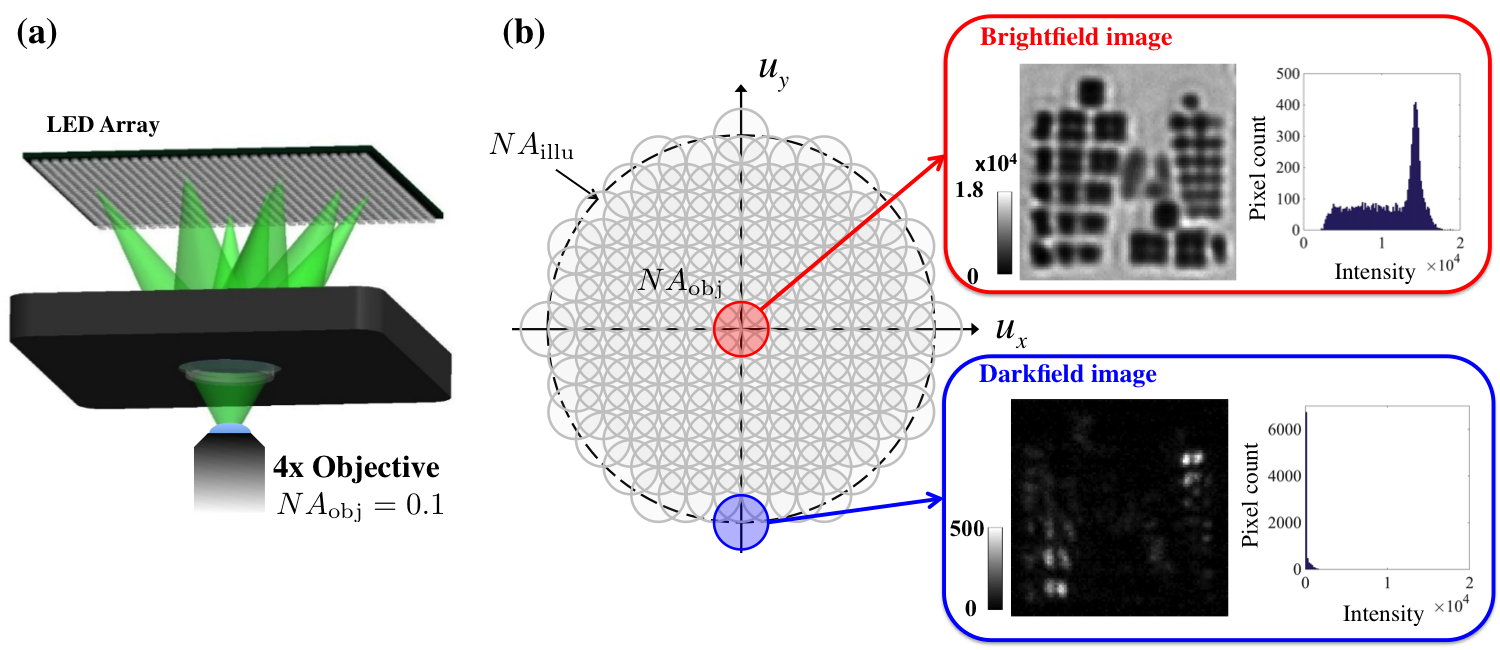}
\vspace{-3mm}
\caption{(a) Experimental setup for Fourier ptychography with an LED array microscope. (b) The sample's Fourier space is synthetically enlarged by capturing multiple images from different illumination angles. Each circle represents the spatial frequency coverage of the image captured by single-LED illumination. Brightfield images have orders of magnitude higher intensity than darkfield (see histograms), resulting in different noise levels.}
\label{Fig5}
\end{figure}

We will show that in order for a phase retrieval scheme to be robust to experimental imperfections, the choice of cost function is of crucial importance. One source of error in our experimental data is measurement noise, including Gaussian noise or Poisson shot noise. Another main source of error is model mis-match, caused by experimental imperfections such as aberrations and LED misalignment. A particular problem of FPM datasets is that they contain both brightfield and darkfield images, which have drastically different intensity levels (see Fig.~\ref{Fig5}). Brightfield images can have several orders of magnitude higher intensity than darkfield images; thus, the amount of Poisson noise will also be significantly higher. If this difference in the noise levels is not properly accounted for, the brightfield noise may drown out the darkfield signal. We will further show that aberrations and LED mis-calibration - the two main model mis-match errors in our experiments - result also in intensity-dependent errors. Thus, by carefully designing the the cost function, we can develop algorithms that are significantly more robust to \textit{both} noise and model mis-match. 

\begin{figure}[H]
\centering
\vspace{-3mm}
\includegraphics[width=12.5cm]{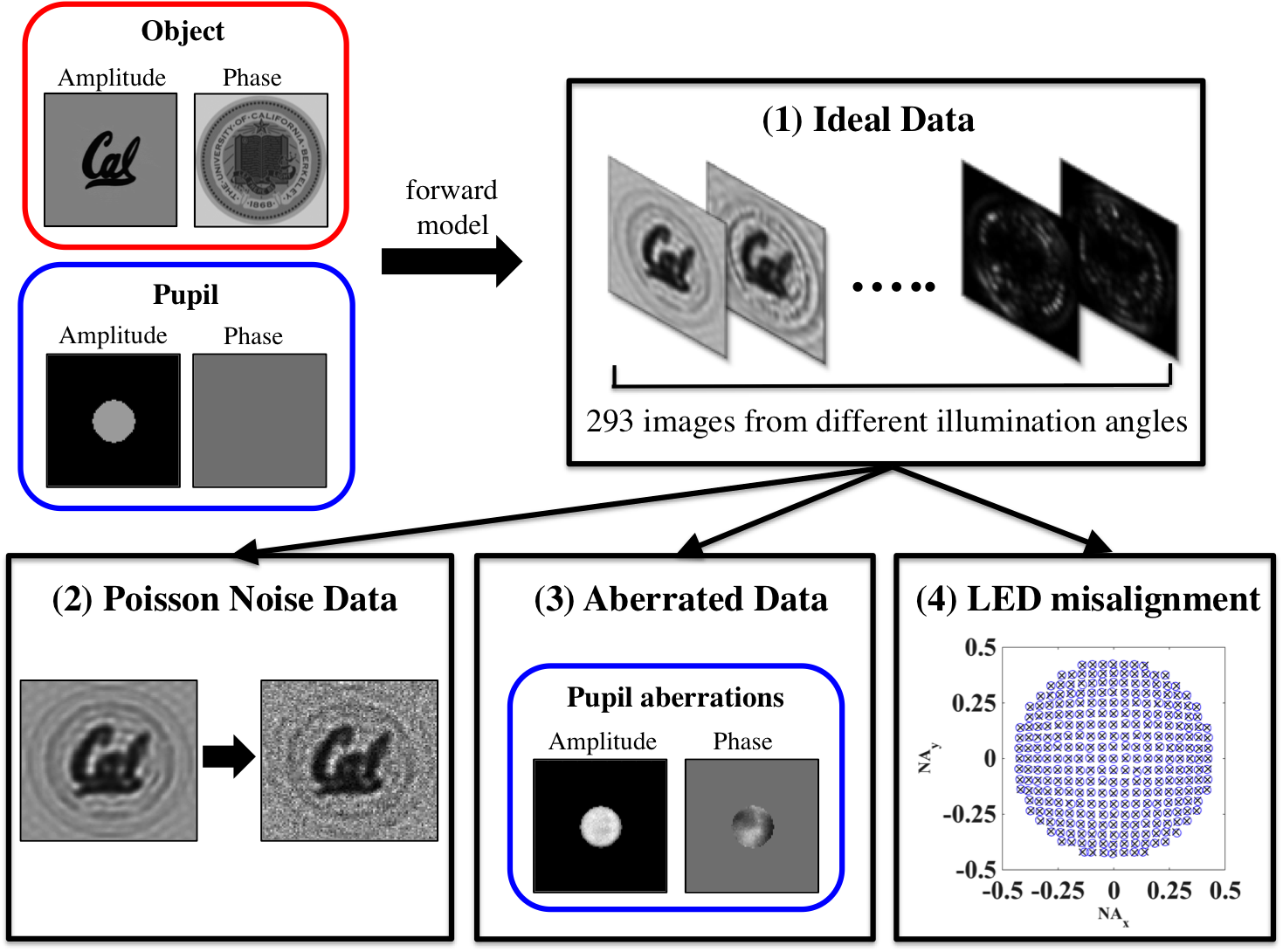}
\vspace{-2mm}
\caption{To explain the artifacts in our experimental results, as well as evaluate the robustness of various algorithms under common types of errors, we simulate several FPM datasets with different types of known error: (1) Ideal data, (2) Poisson noise data, (3) aberrated data, (4) LED misaligned data ($\times$: original position, $\circ$: perturbed position).}
\vspace{-3mm}
\label{Fig1}
\end{figure}

We develop a maximum likelihood theory which provides a flexible framework for formulating the FPM optimization problem with various noise models. In particular, we will focus on Gaussian and Poisson noise models. We find that \textit{amplitude-based} algorithms effectively use a Poisson noise model, while \textit{intensity-based} algorithms use a Gaussian noise model. To illustrate, we simulate four FPM datasets, three of which are contaminated with measurement errors (see Fig.~\ref{Fig1}): Poisson noise, aberrations, and LED misalignment. We compare the performance of various algorithms on these datasets to demonstrate that the imperfections in our experimental data are more consistent with a Poisson noise model. This explains our observations that \textit{amplitude-based} algorithms are more experimentally robust than \textit{intensity-based} algorithms. 

\section{Algorithm Formulation} 

\subsection{Forward problem for Fourier ptychography}

Consider a thin sample with transmission function $o(\mathbf{r})$, where $\mathbf{r} = (x,y)$ represents the 2D spatial coordinates in the sample plane. Assuming that the LED array is sufficiently far from the sample, each LED will illuminate the sample by a plane wave from a different angle, defined by $\exp(i2 \pi \mathbf{u}_\ell\cdot\mathbf{r})$, where $\mathbf{u}_\ell = (u_{\ell,x},u_{\ell,y})$ is the spatial frequency corresponding to the $\ell$-th LED, $\ell = 1, \ldots, N_{img}$. After passing through the sample, the exit wave is the product of the sample and illumination complex fields, $o(\mathbf{r})\exp(i2 \pi \mathbf{u}_\ell\cdot\mathbf{r})$. The tilted plane wave illumination means that the Fourier transform of this exit wave is just a shifted version of the Fourier spectrum of the object, $O(\mathbf{u}-\mathbf{u}_\ell)$, where $O(\mathbf{u}) = \mathcal{F}\{o(\mathbf{r})\}$ and $\mathcal{F}$ is the 2D Fourier transform.  This exit wave then passes through the objective lens, where it is low-pass filtered by the pupil function, $P(\mathbf{u})$, which is usually a circle with its size defined by $NA_{\mathrm{obj}}$. Finally, with $\mathcal{F}^{-1}$ being the 2D inverse Fourier transform, we can write the intensity at the image plane as~\cite{Tian2014}
\begin{equation}
I_\ell(\mathbf{r}) = |\mathcal{F}^{-1}\{ P(\mathbf{u})O(\mathbf{u}-\mathbf{u}_\ell) \}|^2.
\label{eqn_forward}
\end{equation} 

\subsection{Possible noise and simulated dataset} \label{subsec:datasets}

Ideally, all algorithms based on the forward model above should give good reconstructions. However, noise and model mis-match errors cause deviations from our forward model. Thus, a noise model that accurately describes the error will be important for noise tolerance. Heuristically, we have identified three experimental non-idealities that cause error: Poisson noise, aberrations and LED mis-alignment. We aim to separate and analyze the artifacts caused by each through controlled simulations that incur only one type of error. 

The simulated data (Fig.~\ref{Fig1}) uses the same parameters as our experimental setup, where a $32\times 32$ green LED array (central wavelength $\lambda = 514$ nm) is placed 77 mm above the sample. LEDs are nominally 4 mm apart from each other and only the central 293 LEDs are used, giving a maximum $NA_{illu} = 0.45$. Samples are imaged with a $4\times$ objective lens having $NA_{obj} = 0.1$.  

Using our forward model, we simulate four datasets: 
\begin{enumerate}
\itemsep0em
\item Ideal data: no noise is added. The object and pupil follow exactly the FPM forward model that is assumed in the algorithm.
\item Poisson noise data: the ideal data is corrupted by Poisson-distributed noise at each pixel. To emphasize the effect and to emulate experiments with lower-performance sensors, we simulate 20$\times$ more noise than is present in our experiments (details in Section~\ref{sec:model}). 
\item Aberrated data: simulated images are corrupted by imaging system aberrations, which are described by the aberrated complex pupil function shown in Fig.~\ref{Fig1}. The pupil function used in these simulations was obtained from experimental measurements.
\item LED mis-aligned data: the illumination angle of each LED is perturbed slightly (following a normal distribution with standard deviation $\sigma_\theta = 0.2^\circ$). The black $\times$ and blue $\circ$ in Fig.~\ref{Fig1} show the original and perturbed LED positions, respectively. 
\end{enumerate}

To deal with these experimental errors, in the next section we will discuss different noise models for formulating the FPM optimization problem.

\subsection{Optimization problem based on different noise models}\label{sec:model}

Most algorithms solve the FPM problem by minimizing the difference between the measured and estimated amplitude (or intensity), without assuming a noise model. Hence, the FPM problem can be formulated as the following optimization
\begin{equation}
\min_{O(\mathbf{u})}f_A(O(\mathbf{u})) = \min_{O(\mathbf{u})}\sum_\ell \sum_\mathbf{r} |\sqrt{I_\ell(\mathbf{r})} - |\mathcal{F}^{-1}\{P(\mathbf{u}) O(\mathbf{u}-\mathbf{u}_\ell) \}||^2.
\label{eqn_amplitude}
\end{equation}
Since the cost function here, $f_A(O(\mathbf{u}))$, aims to minimize the difference between the estimated amplitude and the measured amplitude, this is the {\it amplitude-based} cost function. By optimizing this cost function, the projection-based algorithms for Fourier ptychography can be obtained~\cite{Zheng:2013gq,Ou:14,Tian2014}, which treat each measurement as an amplitude-based sub-optimization problem. The formulation is used in the traditional Gerchberg-Saxton phase retrieval approach.

If we have information about the statistics of the noise, we can use it in our optimization formulation via the maximum likelihood estimation framework~\cite{thibault2012maximum}. If we assume that our measured images suffer only from white Gaussian noise, then the probability of capturing the measured intensity $I_\ell(\mathbf{r})$ at each pixel, given the estimate of $O(\mathbf{u})$, can be expressed as 
\begin{equation}
p[I_\ell(\mathbf{r})|O(\mathbf{u})] = \frac{1}{\sqrt{2\pi\sigma_w^2}} \exp \left[-\frac{(I_\ell(\mathbf{r}) - \hat{I}_\ell(\mathbf{r}))^2}{2\sigma_w^2}\right],
\label{eqn_Gaussian_prob}
\end{equation}
where $\hat{I}_\ell(\mathbf{r}) = |\mathcal{F}^{-1}\{ P(\mathbf{u})O(\mathbf{u}-\mathbf{u}_\ell) \}|^2$ and $\sigma_w$ is the standard deviation of the Gaussian noise. $\hat{I}_\ell(\mathbf{r})$ and $I_\ell(\mathbf{r})$ denote the estimated and measured intensity, respectively. 

The likelihood function is the overall probability due to all the pixels in all the images and can be calculated as $\prod_\ell \prod_\mathbf{r} p[I_\ell(\mathbf{r})|O(\mathbf{u})]$, assuming measurements from all pixels are independent. In maximum likelihood estimation, the goal is to maximize the likelihood function. However, it is easier to solve this problem by turning the likelihood function into a negative log-likelihood function which can be minimized. The negative log-likelihood function associated with this probability distribution can be calculated as
\begin{eqnarray}
&&\hspace{-0.3 in}\mathcal{L}_{\rm Gaussian} (O(\mathbf{u})) = -\log \prod_\ell \prod_\mathbf{r} p[I_\ell(\mathbf{r})|O(\mathbf{u})] \nonumber \\
&& = \sum_\ell \sum_\mathbf{r} \left[\frac{1}{2}\log (2\pi\sigma_w^2) + \frac{(I_\ell(\mathbf{r}) - \hat{I}_\ell(\mathbf{r}))^2}{2\sigma_w^2} \right].
\label{eqn_Gaussian_log}
\end{eqnarray}

The next step is to minimize this negative log-likelihood function by estimating $O(\mathbf{u})$ so that the overall probability is maximized. For white Gaussian noise, it is assumed that $\sigma_w^2$ are the same across all pixels for all images (i.e. all measurements have the same amount of noise), though this will \emph{not} be the case for FPM datasets. By making a Gaussian noise assumption, the first term in (\ref{eqn_Gaussian_log}) is a constant and can be ignored. The optimization problem then reduces to
\begin{equation}
\min_{O(\mathbf{u})}f_I(O(\mathbf{u})) = \min_{O(\mathbf{u})}\sum_\ell \sum_\mathbf{r} |I_\ell(\mathbf{r}) - |\mathcal{F}^{-1}\{P(\mathbf{u}) O(\mathbf{u}-\mathbf{u}_\ell) \}|^2|^2.
\label{eqn_intensity}
\end{equation}
We call this cost function, $f_I(O(\mathbf{u}))$, the {\it intensity-based} cost function because it aims to minimize the difference between the estimated intensity and the measured intensity. It also implies that noise from each pixel is treated the same and independent of the measured intensity. It will be shown later that the previous implementations of PhaseLift~\cite{horstmeyer2014solving} and Wirtinger flow algorithms~\cite{bian2015fourier} for FPM aimed to optimize this {\it intensity-based} cost function. However, both can be implemented instead with a Poisson likelihood cost function.

If we assume instead that our measured images suffer from Poisson shot noise, then the probability of the measured intensity, $I_\ell(\mathbf{r})$, given the estimate of $O(\mathbf{u})$ can be expressed as 
\begin{equation}
p[I_\ell(\mathbf{r})|O(\mathbf{u})] = \frac{[\hat{I}_\ell(\mathbf{r})]^{I_\ell(\mathbf{r})} \exp [-\hat{I}_\ell(\mathbf{r})]}{I_\ell(\mathbf{r})!} \approx \frac{1}{\sqrt{2\pi\sigma_{\ell,\mathbf{r}}^2}} \exp \left[-\frac{(I_\ell(\mathbf{r}) - \hat{I}_\ell(\mathbf{r}))^2}{2\sigma_{\ell,\mathbf{r}}^2}\right].
\label{eqn_Poisson_prob}
\end{equation}
Note that the Poisson distribution is used to describe the statistics of the incoming photons at each pixel, which is a discrete probability distribution. Here, we assume that the intensity is proportional to the photon count, so we can treat the distribution of the intensity as a Poisson distribution. When the expected value of the Poisson distribution is large, then this Poisson distribution will become more like a Gaussian distribution having a standard deviation proportional to the square root of the intensity, $\sigma_{\ell,\mathbf{r}} \approx \sqrt{I_\ell(\mathbf{r})}$, from the central limit theorem. This means that a large measured intensity at a particular pixel will imply large noise at that pixel. In the simulation, we impose Poisson noise on the measured intensity by distributing each pixel value with a Gaussian distribution and setting the standard deviation to $20\sqrt{I_\ell(\mathbf{r})}$. The negative log-likelihood of the Poisson noise model can then be calculated; the optimization problem is formed by minimizing the negative log-likelihood function with estimation of $O(\mathbf{u})$,
\begin{eqnarray}
&&\hspace{-0.5 in}\min_{O(\mathbf{u})}\mathcal{L}_{\rm Poisson} (O(\mathbf{u})) = \min_{O(\mathbf{u})} \sum_\ell \sum_\mathbf{r} (-I_\ell(\mathbf{r})\log[\hat{I}_\ell(\mathbf{r})]+\hat{I}_\ell(\mathbf{r})+\log[I_\ell(\mathbf{r})!]) \nonumber \\
&&\approx \min_{O(\mathbf{u})}\sum_\ell \sum_\mathbf{r} \frac{(I_\ell(\mathbf{r}) - \hat{I}_\ell(\mathbf{r}))^2}{2\sigma_{\ell,\mathbf{r}}^2}.
\label{eqn_Poisson_log}
\end{eqnarray}
This cost function comes from the likelihood function of the Poisson distribution, so we call it the {\it Poisson-likelihood-based} cost function. It implies that the pixels with larger measured intensity are weighted smaller because they suffer from more noise. Since the brightfield images have more large-value pixels, they are assumed to be more noisy and thus are weighted smaller in the cost function. It is shown in the Appendix that the gradient of this cost function (\ref{eqn_grad_Poisson}) is very similar to that of the amplitude-based cost function (\ref{eqn_grad_fA}), which suggests that the amplitude-based cost function deals well with Poisson-like noise or model mis-match.

\subsection{Vectorization Notation}

For multivariate optimization problems such as (\ref{eqn_amplitude}) and (\ref{eqn_intensity}) , it is convenient to reformulate the problem using linear algebra. First, the functions need to be vectorized. Each of the captured images, $I_\ell(\mathbf{r})$, having $m\times m$ pixels, are raster-scanned into vectors, $\mathbf{I}_\ell$, with size $m^2\times 1$. Since the estimated object transmission function will have higher space-bandwidth product than the raw images, the estimated object should have $n\times n$ pixels, where $n>m$. For convenience, we actually solve for the Fourier space of the object, $O(\mathbf{u})$, which is vectorized into a vector $\mathbf{O}$ with size $n^2\times 1$. Before multiplying the pupil function, the Fourier space of the object is downsampled by a $m^2\times n^2$ matrix $\mathbf{Q}_\ell$. The matrix $\mathbf{Q}_\ell$ transforms a $n^2\times 1$ vector into a $m^2\times 1$ vector by selecting values out of the original vector, so the entries of this matrix are either 1 or 0 and each row contains at most one nonzero element. The pupil function $P(\mathbf{u})$ is vectorized into a vector $\mathbf{P}$ with size $m^2\times 1$. The 2D Fourier transform and inverse transform operator are $m^2\times m^2$ matrices defined as $\mathbf{F}$ and $\mathbf{F}^{-1}$. $|\cdot|$, $|\cdot|^2$, $\sqrt{\cdot}$, and $\cdot/\cdot$ are element-wise operators, and the ${\rm diag}(\cdot)$ operator puts the entries of a vector into the diagonal of a matrix. 

The second step is to rewrite the optimization in vector form using the new parameters. First, the forward model (\ref{eqn_forward}) can be vectorized as
\begin{eqnarray}
&&\hat{\mathbf{I}}_\ell = |\mathbf{g}_\ell|^2 = |\mathbf{F}^{-1}{\rm diag}(\mathbf{P})\mathbf{Q}_\ell\mathbf{O}|^2.
\label{eqn_forward_rew}
\end{eqnarray}
The {\it amplitude-based} cost function (\ref{eqn_amplitude}) can be vectorized as 
\begin{equation}
\min_{\mathbf{O}}f_A(\mathbf{O}) = \min_{\mathbf{O}}\sum_\ell (\sqrt{\mathbf{I}_\ell} - |\mathbf{g}_\ell|)^\dagger(\sqrt{\mathbf{I}_\ell} - |\mathbf{g}_\ell|),
\label{eqn_amplitude_rew}
\end{equation}
where the hyperscript $\dagger$ denotes a Hermitian conjugate. 

Likewise, the {\it intensity-based} cost function (\ref{eqn_intensity}) can be vectorized as 
\begin{equation}
\min_{\mathbf{O}}f_I(\mathbf{O}) = \min_{\mathbf{O}}\sum_\ell (\mathbf{I}_\ell - |\mathbf{g}_\ell|^2)^\dagger(\mathbf{I}_\ell - |\mathbf{g}_\ell|^2).
\label{eqn_intensity_rew}
\end{equation}

The Poisson likelihood cost function is more complicated to be expressed in vector form. First, we rewrite $|\mathbf{g}_\ell|^2$ as
\begin{equation}
|\mathbf{g}_\ell|^2 = {\rm diag} (\bar{\mathbf{g}}_\ell) \mathbf{F}^{-1}{\rm diag} (\mathbf{P})\mathbf{Q}_\ell \mathbf{O} = \mathbf{A}_\ell \mathbf{O} = \begin{bmatrix}
\mathbf{a}_{\ell,1}^\dagger \\
\vdots \\
\mathbf{a}_{\ell,m^2}^\dagger
\end{bmatrix}\mathbf{O},
\label{eqn_Poisson_zL}
\end{equation}
where $\mathbf{A}_\ell = {\rm diag} (\bar{\mathbf{g}}_\ell) \mathbf{F}^{-1}{\rm diag} (\mathbf{P})\mathbf{Q}_\ell$ is a $m^2 \times n^2$ matrix with $m^2 \times 1$ row vectors, $\mathbf{a}_{\ell,j}^\dagger$, $j = 1, \ldots, m^2$, and $\bar{\mathbf{g}}_\ell$ denotes the complex conjugate of vector $\mathbf{g}_\ell$. 
Then the likelihood function can be rewritten as
\begin{equation}
\min_{\mathbf{O}}\mathcal{L}_{\rm Poisson} (\mathbf{O}) = \sum_\ell \sum_j [-I_{\ell,j}\log(\mathbf{a}_{\ell,j}^\dagger\mathbf{O}) + \mathbf{a}_{\ell,j}^\dagger \mathbf{O} + \log (I_{\ell,j}!)].
\label{eqn_Poisson_rew}
\end{equation}

To minimize (\ref{eqn_amplitude_rew}), (\ref{eqn_intensity_rew}) or (\ref{eqn_Poisson_rew}) using an iterative optimization algorithm, the gradients (and possibly Hessians) of the cost functions need to be calculated, both of which are shown in the Appendix. Since (\ref{eqn_amplitude_rew}), (\ref{eqn_intensity_rew}) and (\ref{eqn_Poisson_rew}) are all real-valued functions of a complex vector $\mathbf{O}$, that means that $\mathbf{O}$ and $\bar{\mathbf{O}}$ should be treated independently in the derivative calculation, which is based on the CR-calculus discussed in~\cite{Ken2009} and the similar formulation for traditional ptychography discussed in~\cite{Yang:2011ty}.

\section{Derivation of algorithms}\label{sec:Algos}

The basic formulation of the optimization problem in FPM has been described in the last section and the derivative calculation has been done in the Appendix, so we now turn our attention to describing how each algorithm solves the optimization problem based on different cost functions. The key step will be in how each algorithm updates the estimate of the object at each iteration. We compare the existing algorithms for FPM and also implement a new second-order global Newton's method under different cost functions, for comparison. The initialization for all algorithms is the same - the amplitude of the image from the on-axis LED illumination.

\subsection{First-order methods}

\subsubsection{Sequential gradient descent (Gerchberg-Saxton)~\cite{Zheng:2013gq,Ou:14}}

For the implementation in~\cite{Zheng:2013gq,Ou:14}, the algorithm aims to optimize the amplitude-based cost function (\ref{eqn_amplitude_rew}). It is the simplest to implement and, in this case, equivalent to the Gerchberg-Saxton approach of simply replacing known information in real and Fourier space. Since the sequential strategy treats a single image as an optimization problem, the cost function for each problem is just one component of Eq.~\eqref{eqn_amplitude_rew} and is defined as 
\begin{equation}
f_{A,\ell} (\mathbf{O}) = (\sqrt{\mathbf{I}_\ell} - |\mathbf{g}_\ell|)^\dagger (\sqrt{\mathbf{I}_\ell} - |\mathbf{g}_\ell|),
\label{eqn_amplitude_seq}
\end{equation}
where $\ell$ denotes the index of each measurement. 

The derivative of this cost function is thus a component of Eq.~\eqref{eqn_grad_fA} and can be expressed as
\begin{eqnarray}
&&\nabla_{\mathbf{O}} f_{A,\ell} (\mathbf{O}) = -\mathbf{Q}_\ell^\dagger {\rm diag}(\bar{\mathbf{P}}) \left[ \mathbf{F} {\rm diag}\left( \frac{\sqrt{\mathbf{I}_\ell}}{|\mathbf{g}_\ell|}\right) \mathbf{g}_\ell - {\rm diag}(\mathbf{P}) \mathbf{Q}_{\ell} \mathbf{O} \right].
\label{eqn_amplitude_seq_grad}
\end{eqnarray}

The update equation for this sequential amplitude-based algorithm is then a gradient descent with the descent direction given by Eq.~\eqref{eqn_amplitude_seq_grad} and step size $1/|\mathbf{P}|^2_{\rm max}$: 
\begin{equation}
\mathbf{O}^{(i,\ell+1)} = \mathbf{O}^{(i,\ell)} - \frac{1}{|\mathbf{P}|^2_{\rm max}}\nabla_{\mathbf{O}} f_{A,\ell+1} (\mathbf{O}^{(i,\ell)}),
\label{eqn_amplitude_seq_update}
\end{equation}
where $i$ indicates the iteration number, which goes to $i+1$ after running through all the measurements from $\ell = 1$ to $\ell = N_{img}$. This algorithm adopts the alternating projection phase retrieval approach. The first projection in the real domain is the amplitude replacement operation ${\rm diag}\left( \frac{\sqrt{\mathbf{I}_\ell}}{|\mathbf{g}_\ell|}\right)\mathbf{g}_\ell$, and the second projection is to project the previous estimated Fourier region ${\rm diag}(\mathbf{P}) \mathbf{Q}_{\ell} \mathbf{O}$ onto the updated Fourier region $\mathbf{F} {\rm diag}\left( \frac{\sqrt{\mathbf{I}_\ell}}{|\mathbf{g}_\ell|}\right) \mathbf{g}_\ell$. 

It is worth noting that the algorithm in~\cite{Zheng:2013gq} directly replaces $\mathbf{F} {\rm diag}\left( \frac{\sqrt{\mathbf{I}_\ell}}{|\mathbf{g}_\ell|}\right) \mathbf{g}_\ell$ in the Fourier domain at each sub-iteration. A similar algorithm in~\cite{Ou:14}, introduced for simultaneous aberration recovery, has the same form as Eq. \eqref{eqn_amplitude_seq_update} that implements gradient descent in the Fourier domain. However, when there is no pupil estimation, then  $\mathbf{P}$ becomes a pure support function with one inside the support and zero outside. In this situation, these two algorithms become exactly the same, and thus we refer to both as sequential gradient descent or Gerchberg-Saxton algorithm. 

\subsubsection{Wirtinger flow algorithm~\cite{bian2015fourier,Candes:2014ur}}

The Wirtinger flow optimization framework was originally proposed to iteratively solve the coded-mask phase retrieval problem using nonlinear optimization~\cite{Candes:2014ur}. It is a gradient descent method implemented with a special initialization and special step sizes. For the FPM implementation described in~\cite{bian2015fourier}, the \textit{intensity-based} cost function is used. Thus, the update equation for the object transmission function $\mathbf{O}$ can be expressed as
\begin{equation}
\mathbf{O}^{(i+1)} = \mathbf{O}^{(i)} - \alpha^{(i)} \nabla_{\mathbf{O}} f_I(\mathbf{O}^{(i)}),
\label{eqn_grad_des_wir}
\end{equation}
where the step size is calculated by
\begin{equation}
\alpha^{(i)} = \frac{\min(1-e^{-i/i_0},\theta_{\rm max})}{(\mathbf{O}^{(0)})^\dagger\mathbf{O}^{(0)}},
\end{equation}
where $\nabla_{\mathbf{O}} f_I(\mathbf{O}^{(i)})$ is the gradient of the intensity-based cost function calculated in (\ref{eqn_grad_fI}), and $i_0$ and $\theta_{\rm max}$ are user-chosen parameters to calculate the step size. 

In the previously proposed FPM implementation of Wirtinger flow~\cite{bian2015fourier}, the algorithm deviates somewhat from the original theory proposed in~\cite{Candes:2014ur}. First, there is an additional term in the cost function to deal with additive noise. Second, the initialization used in~\cite{bian2015fourier} is not the proposed one in~\cite{Candes:2014ur}, but rather a low-resolution captured image. So the algorithm in~\cite{bian2015fourier} is essentially a gradient descent method with the special step size based on the intensity-based cost function and is not guaranteed to converge to the global minimum. 

The Wirtinger flow algorithm can be implemented with different cost functions simply by replacing the original \textit{intensity-based} gradient with the other gradients derived in the Appendix. For comparison, we have implemented the Wirtinger flow algorithm using all three of the cost functions described here: \textit{amplitude-based}, \textit{intensity-based} and \textit{Poisson-likelihood-based}. The results are compared in Fig.~\ref{Fig4} with experimental data and Section~\ref{sec:comparison} with simulated data.

\subsection{Second-order methods}

Beyond first-order, a second-order optimization method can improve the convergence speed and stability of the algorithm, especially for nonlinear and non-convex problems. Second-order methods (e.g. Newton's method) use both the first and second derivatives (Hessian) of the cost function to create a better update at each iteration. As a result, they generally require fewer iterations and move more directly towards the solution. The difficulty of second-order implementations is in computing the Hessian matrix, whose size scales quadratically with the size of the image. As a result, approximations to the Hessian are often used (known as \textit{quasi-Newton} methods) to trade performance for computational efficiency.

\subsubsection{Sequential Gauss-Newton method~\cite{Tian2014}}

First, we look at a Gauss-Newton method based on the \textit{amplitude-based} cost function, which approximates the Hessian matrix as a multiplication of its Jacobian matrix:
\begin{eqnarray}
&&\mathbf{H}^A_{\mathbf{c}\mathbf{c},\ell} \approx \left(\frac{\partial\mathbf{f}_{A\ell}}{\partial\mathbf{c}}\right)^\dagger \left(\frac{\partial\mathbf{f}_{A\ell}}{\partial\mathbf{c}}\right) \nonumber \\
&&\hspace{0.3 in}=\begin{bmatrix}
\frac{1}{2}\mathbf{Q}_\ell^\dagger {\rm diag}(|\mathbf{P}|^2)\mathbf{Q}_\ell &
\mathbf{Q}_\ell^\dagger {\rm diag}(\bar{\mathbf{P}})\mathbf{F} {\rm diag}\left(\frac{\mathbf{g}_\ell^2}{|\mathbf{g}_\ell|^2}\right) \bar{\mathbf{F}}^{-1} {\rm diag}(\bar{\mathbf{P}})\bar{\mathbf{Q}}_\ell \\
\mathbf{Q}_\ell^T {\rm diag}(\mathbf{P})\bar{\mathbf{F}} {\rm diag}\left(\frac{\bar{\mathbf{g}}_\ell^2}{|\mathbf{g}_\ell|^2}\right) \mathbf{F}^{-1} {\rm diag}(\mathbf{P})\mathbf{Q}_\ell &
\frac{1}{2}\mathbf{Q}_\ell^T {\rm diag}(|\mathbf{P}|^2)\bar{\mathbf{Q}}_\ell
\end{bmatrix}, \nonumber \\
&&
\label{eqn_amplitude_seq_hessian_approx1}
\end{eqnarray}
where $\mathbf{c} = (\mathbf{O}^T,\bar{\mathbf{O}}^T)^T$ (See Appendix).
Since the inversion of this Hessian matrix requires very high computational cost, we approximate the Hessian by dropping all the off-diagonal terms of the Hessian matrix. Further, the inversion of the Hessian matrix may be an ill-posed problem, so a constant regularizer is adopted. In the end, the approximated Hessian inversion becomes 
\begin{eqnarray}
(\mathbf{H}^A_{\mathbf{c}\mathbf{c},\ell})^{-1} \approx \begin{bmatrix}
2\mathbf{Q}_\ell^\dagger {\rm diag}\left(\frac{1}{|\mathbf{P}|^2+\Delta}\right) \mathbf{Q}_\ell &
\mathbf{0} \\
\mathbf{0} &
2\mathbf{Q}_\ell^T {\rm diag}\left(\frac{1}{|\mathbf{P}|^2+\Delta}\right) \bar{\mathbf{Q}}_\ell
\end{bmatrix},
\label{eqn_amplitude_seq_hessian_approx2}
\end{eqnarray}
where $\Delta$ is a constant vector with all the entries equal to a constant regularizer $\delta$ over all pixels. 

By applying Newton's update, Eq.~\eqref{eqn_Newton}, with this approximated Hessian inversion, the new estimate of $\mathbf{O}$ can be expressed as
\begin{eqnarray}
\begin{bmatrix}
\mathbf{O}^{(i,\ell+1)} \\
\bar{\mathbf{O}}^{(i,\ell+1)}
\end{bmatrix}
=
\begin{bmatrix}
\mathbf{O}^{(i,\ell)} \\
\bar{\mathbf{O}}^{(i,\ell)}
\end{bmatrix}
- 
\begin{bmatrix}
\mathbf{Q}^{\dagger}_\ell {\rm diag} \left(\frac{|\mathbf{P}|}{|\mathbf{P}|_{\rm max}}\right)\mathbf{Q}_\ell & \mathbf{0} \\
\mathbf{0} &\mathbf{Q}^{T}_\ell {\rm diag}\left(\frac{|\mathbf{P}|}{|\mathbf{P}|_{\rm max}}\right) \bar{\mathbf{Q}}_\ell
\end{bmatrix}
(\mathbf{H}^A_{\mathbf{c}\mathbf{c},\ell})^{-1} 
\begin{bmatrix}
\nabla_{\mathbf{O}} f_{A,\ell+1} (\mathbf{O}^{(i,\ell)}) \\
\nabla_{\bar{\mathbf{O}}} f_{A,\ell+1} (\mathbf{O}^{(i,\ell)}) 
\end{bmatrix},
\label{eqn_amplitude_seq_hessian_update}
\end{eqnarray}
where the ${\rm diag}\left(\frac{|\mathbf{P}|}{|\mathbf{P}|_{\rm max}}\right)$ part is the step size for this descent direction. Note that when $\mathbf{P}$ is a constant having either 0 or 1 values, this method is reduced to the sequential gradient descent method with a tunable regularizer $\delta$. In practice, however, we also simultaneously update $\mathbf{P}$ (see Section~\ref{sec: Pupil}), so the second-order optimization procedure becomes more crucial. 

\subsubsection{New algorithm implementing a global Newton's method} \label{sec: Newton}

Since we expect second-order methods to perform better than first-order, and also global methods to be more stable than sequential, we propose a new global second-order (Newton's) method, and show the results compared against other methods. For completeness, we implement all three of amplitude, intensity, and Poisson-likelihood-based cost functions, showing that the amplitude and Poisson-likelihood-based cost functions indeed perform better. The difficult step in deriving a Newton's method for this problem is in calculating the gradients and Hessians of the cost functions directly, without approximations. In the Appendix, we show our derivation, and in this section we use the results with a typical Newton's update equation:

\begin{eqnarray}
\begin{bmatrix}
\mathbf{O}^{(i+1)} \\
\bar{\mathbf{O}}^{(i+1)}
\end{bmatrix} = 
\begin{bmatrix}
\mathbf{O}^{(i)} \\
\bar{\mathbf{O}}^{(i)}
\end{bmatrix} - \alpha^{(i)} (\mathbf{H}_{\mathbf{c}\mathbf{c}})^{-1} 
\begin{bmatrix}
\nabla_{\mathbf{O}}f(\mathbf{O}^{(i)}) \\
\nabla_{\bar{\mathbf{O}}}f(\mathbf{O}^{(i)})
\end{bmatrix}.
\label{eqn_Newton_general}
\end{eqnarray}
The inverse of the Hessian matrix, $(\mathbf{H}_{\mathbf{c}\mathbf{c}})^{-1}$, is solved efficiently by a conjugate gradient matrix inversion iterative solver as described in~\cite{Nocedal2006}. $\alpha^{(i)}$ is determined by the backtracking line search algorithm at each iteration, as described in~\cite{Nocedal:2006uv}.
The exact form of the cost function and the Hessian depends on the algorithm used. For \textit{amplitude-based} Newton's algorithm, $f(\mathbf{O}) = f_A(\mathbf{O})$ and $\mathbf{H}_{\mathbf{c}\mathbf{c}} = \mathbf{H}^A_{\mathbf{c}\mathbf{c}}$; for \textit{intensity-based} Newton's algorithm, $f(\mathbf{O}) = f_I(\mathbf{O})$ and $\mathbf{H}_{\mathbf{c}\mathbf{c}} = \mathbf{H}^I_{\mathbf{c}\mathbf{c}}$; for \textit{Poisson-likelihood-based} Newton's algorithm, $f(\mathbf{O}) = \mathcal{L}_{\rm Gaussian}(\mathbf{O})$ and $\mathbf{H}_{\mathbf{c}\mathbf{c}} = \mathbf{H}^P_{\mathbf{c}\mathbf{c}}$.

\subsection{Convex-based methods}

\subsubsection{PhaseLift algorithm~\cite{horstmeyer2014solving,Recht:2010ht,Candes:2013ka,candes2013phaselift,Burer:2003fg}}

The PhaseLift formulation for phase retrieval is conceptually quite different than the previous methods described here. The idea is to lift the non-convex problem into a higher-dimensional space in which it is convex, thereby guaranteeing convergence to the global solution. To do this, the cost function of $\mathbf{O}$ is reformulated into that of a rank-1 matrix $\mathbf{X} = \mathbf{O}\mathbf{O}^\dagger$ and the goal is to estimate $\mathbf{X}$ instead of $\mathbf{O}$. The process of reformulation can be expressed as~\cite{horstmeyer2014solving}
\begin{eqnarray}
&&\mathbf{g} = 
\begin{bmatrix}
\mathbf{g}_1 \\
\vdots \\
\mathbf{g}_{N_{img}}
\end{bmatrix}
= 
\begin{bmatrix}
\mathbf{F}^{-1} & \cdots & \mathbf{0} \\
\vdots & \ddots & \vdots \\
\mathbf{0} & \cdots & \mathbf{F}^{-1}
\end{bmatrix}
\begin{bmatrix}
{\rm diag}(\mathbf{P}) & \cdots & \mathbf{0} \\
\vdots & \ddots & \vdots \\
\mathbf{0} & \cdots & {\rm diag}(\mathbf{P})
\end{bmatrix}
\begin{bmatrix}
\mathbf{Q}_1 \\
\vdots \\
\mathbf{Q}_{N_{img}}
\end{bmatrix}
\mathbf{O} \nonumber \\
&&\hspace{0.1 in} = \mathbf{D} \mathbf{O} = 
\begin{bmatrix}
\mathbf{d}_1^\dagger \\
\vdots \\
\mathbf{d}_{N_{img}m^2}^\dagger
\end{bmatrix}
\mathbf{O},
\label{eqn_phaselift_form}
\end{eqnarray}
where $\mathbf{D}$ is an $N_{img}m^2\times N_{img}m^2$ operator combining the inverse Fourier transform, pupil cropping, and the downsampling operation with row vectors denoted by $\mathbf{d}_j^\dagger$. 

Hence, the estimated intensity $|\mathbf{g}|^2$ as a function of $\mathbf{X}$ can be expressed 
\begin{eqnarray}
&&\hspace{-0.6 in}|\mathbf{g}|^2 = 
\begin{bmatrix}
\mathbf{O}^\dagger \mathbf{d}_1 \mathbf{d}_1^\dagger \mathbf{O} \\
\vdots \\
\mathbf{O}^\dagger \mathbf{d}_{N_{img}m^2} \mathbf{d}_{N_{img}m^2}^\dagger \mathbf{O}
\end{bmatrix}
=
\begin{bmatrix}
\mathrm{Tr}(\mathbf{d}_1 \mathbf{d}_1^\dagger \mathbf{O}\mathbf{O}^\dagger) \\
\vdots \\
\mathrm{Tr}(\mathbf{d}_{N_{img}m^2} \mathbf{d}_{N_{img}m^2}^\dagger \mathbf{O}\mathbf{O}^\dagger)
\end{bmatrix}
=
\begin{bmatrix}
\mathrm{Tr}(\mathbf{D}_1 \mathbf{X}) \\
\vdots \\
\mathrm{Tr}(\mathbf{D}_{N_{img}m^2} \mathbf{X})
\end{bmatrix}
= \mathcal{A}(\mathbf{X}),
\label{eqn_phaselift_tr}
\end{eqnarray}
where $\mathcal{A}$ is a linear operator transforming $\mathbf{X}$ into $|\mathbf{g}|^2$. In Section~\ref{sec:model}, we discussed three different cost functions. Only the intensity-based and Poisson-likelihood-based cost functions are convex on the estimated intensity, $\hat{I}_\ell(\mathbf{r})$, which is a component of $\mathcal{A}(\mathbf{X})$. Thus, the intensity-based and Poisson-likelihood-based cost functions can be turned into a convex function on $\mathbf{X}$ through this transformation. For the implementation in~\cite{horstmeyer2014solving}, by defining $\mathbf{I} = [\mathbf{I}_1^T,\ldots,\mathbf{I}_{N_{img}}^T]^T$, the intensity-based cost function can be expressed as
\begin{eqnarray}
&&f_I (\mathbf{X}) = (\mathbf{I} - |\mathbf{g}|^2)^\dagger(\mathbf{I} - |\mathbf{g}|^2) \nonumber \\
&&\hspace{0.3 in}= (\mathbf{I} - \mathcal{A} (\mathbf{X}))^\dagger(\mathbf{I} - \mathcal{A} (\mathbf{X})).
\label{eqn_phaselift_cost}
\end{eqnarray}

Since $\mathbf{X}$ is a rank-1 matrix, we then minimize the rank of $\mathbf{X}$ subject to $\mathbf{I} = \mathcal{A} (\mathbf{X})$. However, the rank minimization problem is NP-hard. Therefore, a convex relaxation~\cite{Recht:2010ht,Candes:2013ka,candes2013phaselift} is used instead to transform the problem into a trace minimization problem. Under this relaxation, the optimization problem becomes
\begin{equation}
\min_{\mathbf{X}}f'_I(\mathbf{X}) = \min_{\mathbf{X}}(\mathbf{I} - \mathcal{A} (\mathbf{X}))^\dagger(\mathbf{I} - \mathcal{A} (\mathbf{X}))  + \alpha \mathrm{Tr}(\mathbf{X}),
\label{eqn_phaselift_cost2}
\end{equation}
where $\alpha$ is a regularization variable that depends on the noise level.

The problem with this new approach is that by increasing the dimensionality of the problem, the size of the matrix $\mathbf{X}$ has become $n^2\times n^2$, which is too large to store and calculate eigenvalue decomposition on a normal computer. To avoid these computational problems, we do not directly solve (\ref{eqn_phaselift_cost2}), but rather apply a factorization to $\mathbf{X} = \mathbf{R}\mathbf{R}^\dagger$, where $\mathbf{R}$ is an $n^2 \times k$ matrix. $\mathbf{X}$ is a rank-1 matrix so $k$ is set to be 1 ($\mathbf{R}$ becomes $\mathbf{O}$). This new problem is then solved effectively using the augmented Lagrangian multiplier, by modifying the original cost function~\cite{horstmeyer2014solving,Burer:2003fg}
\begin{equation}
\min_{\mathbf{R}}f_{AL,I}(\mathbf{R}) = \min_{\mathbf{R}}\frac{\sigma}{2}(\mathbf{I} - \mathcal{A} (\mathbf{R}\mathbf{R}^\dagger))^\dagger(\mathbf{I} - \mathcal{A} (\mathbf{R}\mathbf{R}^\dagger))  + \mathbf{y}^T (\mathbf{I} - \mathcal{A} (\mathbf{R}\mathbf{R}^\dagger)) + \mathrm{Tr}(\mathbf{R}\mathbf{R}^\dagger),
\label{eqn_phaselift_factor}
\end{equation}
where $\mathbf{y}$, $N_{img}m^2\times 1$ vector, is the Lagrangian multiplier, and $\sigma \ge 0$ is the augmented Lagrangian multiplier. Both are parameters that can be tuned to give a better reconstruction. By taking the derivative of this cost function with respect to $\mathbf{R}$ and updating $\mathbf{R}$ in each iteration, the optimization problem can then be solved~\cite{Burer:2003fg}. Unfortunately, after these modifications, the problem becomes non-convex because of the minimization with respect to $\mathbf{R}$ instead of $\mathbf{X}$, and thus is no longer provable. 

In order to provide a more familiar form for comparing the PhaseLift algorithm to the others discussed in this paper, we define $\mathbf{y} = [\mathbf{y}_1^T, \ldots, \mathbf{y}_{N_{img}}^T]^T$, where $\mathbf{y}_i$ is $m^2\times 1$ vector, so that the minimization problem in Eq.~\eqref{eqn_phaselift_factor} becomes
\begin{equation}
\min_{\mathbf{O}}f_{AL,I}(\mathbf{O}) = \min_{\mathbf{O}} \frac{\sigma}{2} \left[\sum_{\ell} (\mathbf{I}_\ell - |\mathbf{g}_\ell|^2 + \frac{2}{\sigma}\mathbf{y}_\ell)^\dagger (\mathbf{I}_\ell - |\mathbf{g}_\ell|^2)\right]  + \mathbf{O}^\dagger\mathbf{O}.
\label{eqn_phaselift_vector}
\end{equation}
Now, we see that the PhaseLift implementation is essentially an intensity-based cost function with an additional constraint that may deal better with noise.

The corresponding derivative of the cost function is calculated as in the previous section: 
\begin{equation}
\nabla_\mathbf{O} f_{AL,I}(\mathbf{O}) = -\sigma \sum_\ell \mathbf{Q}_\ell^\dagger  {\rm diag}(\bar{\mathbf{P}})  \mathbf{F} {\rm diag}(\mathbf{g}_\ell)\left( \mathbf{I}_\ell - |\mathbf{g}_\ell|^2 +\frac{1}{\sigma}\mathbf{y}_\ell \right) + \mathbf{O}.
\label{eqn_phaselift_derivative}
\end{equation}
When $\sigma$ is large compared to the component of $\mathbf{y}_\ell$ and $\mathbf{O}$, the factorized PhaseLift formulation with rank-1 $\mathbf{X}$ is equivalent to the intensity-based optimization problem discussed in the previous section. To solve this optimization problem, a quasi-Newton algorithm called L-BFGS (Limited-memory Broyden-Fletcher-Goldfarb-Shanno) method~\cite{Nocedal:2006uv}, which is a second-order method using an approximated Hessian inversion from previous gradients, is adopted. 

We note that although the PhaseLift algorithm can also be implemented with the \textit{Poisson-likelihood-based} cost function, the algorithm in the rank-1 case is equivalent to our global Newton's method discussed in Section~\ref{sec: Newton} for the same reason as in the above analysis.

\section{Performance analysis of various algorithms}\label{sec:comparison}

In this section, we compare the algorithms described in Section~\ref{sec:Algos} using experimental data, as well as simulated data that mimics the experimental errors described in Section~\ref{subsec:datasets}. We find that second-order optimization generally performs better than first-order, while global methods do not give significant improvement over sequential. Further, we explain why the cost function is a key consideration in choosing an algorithm by explaining the cause of the high-frequency artifacts that result from \textit{intensity-based} algorithms. Interestingly, the two model mis-match errors (aberrations and LED mis-alignment) behave similarly to Poisson noise, in that they also give intensity-dependent errors. Hence, the amplitude and Poisson likelihood algorithms are more robust not only to Poisson noise, but also to model mis-match errors.

\subsection{Reconstruction of the simulated and experimental data}

\begin{figure}[H]
\centering
\vspace{-3mm}
\includegraphics[width=10.5cm]{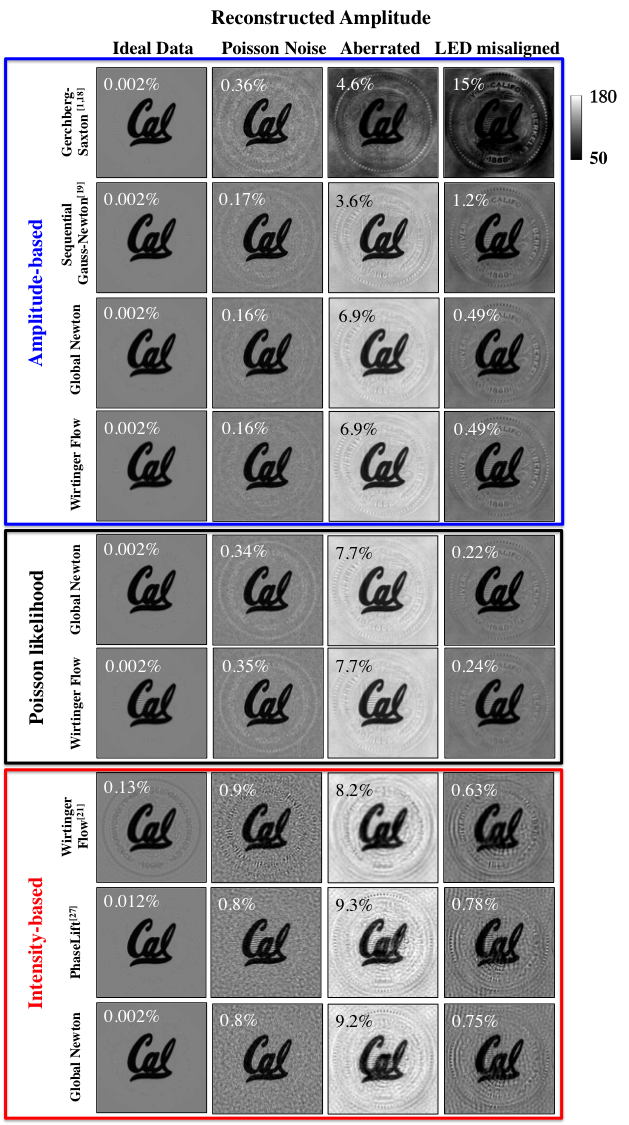}
\vspace{-2mm}
\caption{Reconstructed amplitude from simulated datasets with three types of errors, using different algorithms. The intensity-based algorithms suffer from high frequency artifacts under both noise and model mis-match errors. The percentage on the top left corner of each image is the relative error of each reconstruction.}
\label{Fig2}
\end{figure}

\begin{figure}[H]
\centering
\vspace{-3mm}
\includegraphics[width=10.5cm]{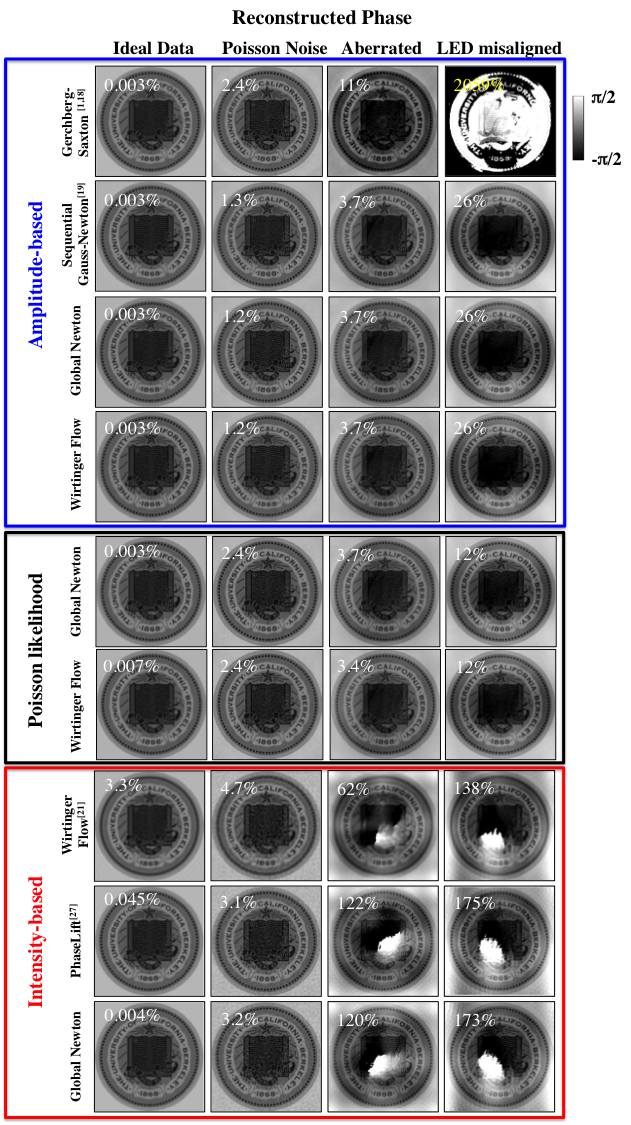}
\caption{Reconstructed phase from simulated datasets with three types of errors, using different algorithms. The intensity-based algorithms suffer from phase wrapping artifacts under both noise and model mis-match errors. The percentage on the top left corner of each image is the relative error of each reconstruction.}
\label{Fig3}
\end{figure}

Next, we use each of the algorithms described in Section~\ref{sec:Algos} to reconstruct amplitude and phase from the datasets simulated in Section~\ref{subsec:datasets}, in order to quantify performance under various experimental error types by comparing against the ground truth input. Figures~\ref{Fig2} and ~\ref{Fig3} show the reconstructed amplitude and phase, respectively. On the top left corner of each image we give the relative error of the reconstruction, defined as 
\begin{equation}
{\rm Error} = \frac{\|\mathbf{O}_{\rm recover} - \mathbf{O}_{\rm true}\|_2^2}{\|\mathbf{O}_{\rm true}\|_2^2},
\label{eqn_relative_error}
\end{equation}
where $\mathbf{O}_{\rm recover}$ and $\mathbf{O}_{\rm true}$ are the reconstructed and true images, respectively, in vector form. In order to ensure that all algorithms converge to their stable solutions, we use 200 iterations for each algorithm, except for Wirtinger flow, which requires 500 iterations. The tuning parameters for each algorithm are summarized in Table~\ref{tab_parameter}. We have attempted to optimize each parameter as fairly as possible; for example, we use a large $\sigma$ in the PhaseLift algorithm to achieve a better reconstruction. Small $\sigma$ trades resolution for flatter background artifacts. 

\begin{table}[h]
\vspace{-3mm}
\caption{Tuning Parameters}
\centering
\begin{tabular}{|c|c|c|c|c|c|}
\hline
\makecell{Gerchberg \\ Saxton} & \makecell{Sequential \\ Gauss-Newton} & \makecell{Amplitude \\ Newton} & \makecell{Amplitude \\ Wirtinger} & \makecell{Poisson \\ Newton} & \makecell{Poisson \\ Wirtinger}  \\ \hline
N/A & $\delta = 5$ & N/A &  \makecell{$i_0 = 10$ \\ $\theta_{\rm max} = 0.05$} & N/A & \makecell{$i_0 = 10$ \\ $\theta_{\rm max} = 0.05$}  \\ \hline  
\makecell{Intensity \\ Wirtinger} & PhaseLift & \makecell{Intensity \\ Newton} & & & \\ \hline
\makecell{$i_0 = 10$ \\ $\theta_{\rm max} = 1$} & $\sigma = 10^{10}$ & N/A & & & \\ \hline
\end{tabular}
\label{tab_parameter}
\end{table}

In analyzing results from the simulated datasets, we find that algorithms with the same cost function give similar reconstruction artifacts. For example, the \textit{intensity-based} algorithms suffer from high-frequency artifacts and phase wrapping when the data is not perfect. Almost all algorithms give a satisfactory reconstruction when using the error-free ideal dataset, except for \textit{intensity-based} Wirtinger flow, which suffers some phase-amplitude leakage and phase blurring (see Figs.~\ref{Fig2}-\ref{Fig3}). When the dataset contains noise or model mis-match, we observe a distinct trend that \textit{amplitude-based} and \textit{Poisson-likelihood-based} algorithms give a better result, compared with \textit{intensity-based} algorithms. The exception to this trend is the Gerchberg-Saxton algorithm, which is somewhat unstable and gets stuck in local minima, so is not robust to any type of error. 

The goal of our simulations was to determine the main error sources that cause artifacts in the experimental reconstructions of Fig.~\ref{Fig4}. Since the experiments contain combined errors from multiple sources, it is difficult to attribute artifacts to any particular type of error. We find, however, that all three of our main error sources cause similar artifacts, hence our experimental results may be corrupted by any of Poisson noise, aberration, or LED misalignment. For example, notice that our simulated error-corrupted data all results in high-frequency artifacts when using \textit{intensity-based} algorithms, similar to the experimental results. The Gerchberg-Saxton result also displays low-frequency errors in simulation, as in experiment. The fact that both noise and model mis-match create similar artifacts is unexpected, since they are very different error mechanisms. We explain below why all three are intensity-dependent errors, which is the reason why the cost function choice is so important for robustness. The consequence is that algorithms which use a more accurate noise model (amplitude and Poisson likelihood-based) will not only be more robust to noise, but also to model mis-match errors.

\begin{figure}[H]
\centering
\hspace{-10mm}
\vspace{-3mm}
\includegraphics[width=9cm]{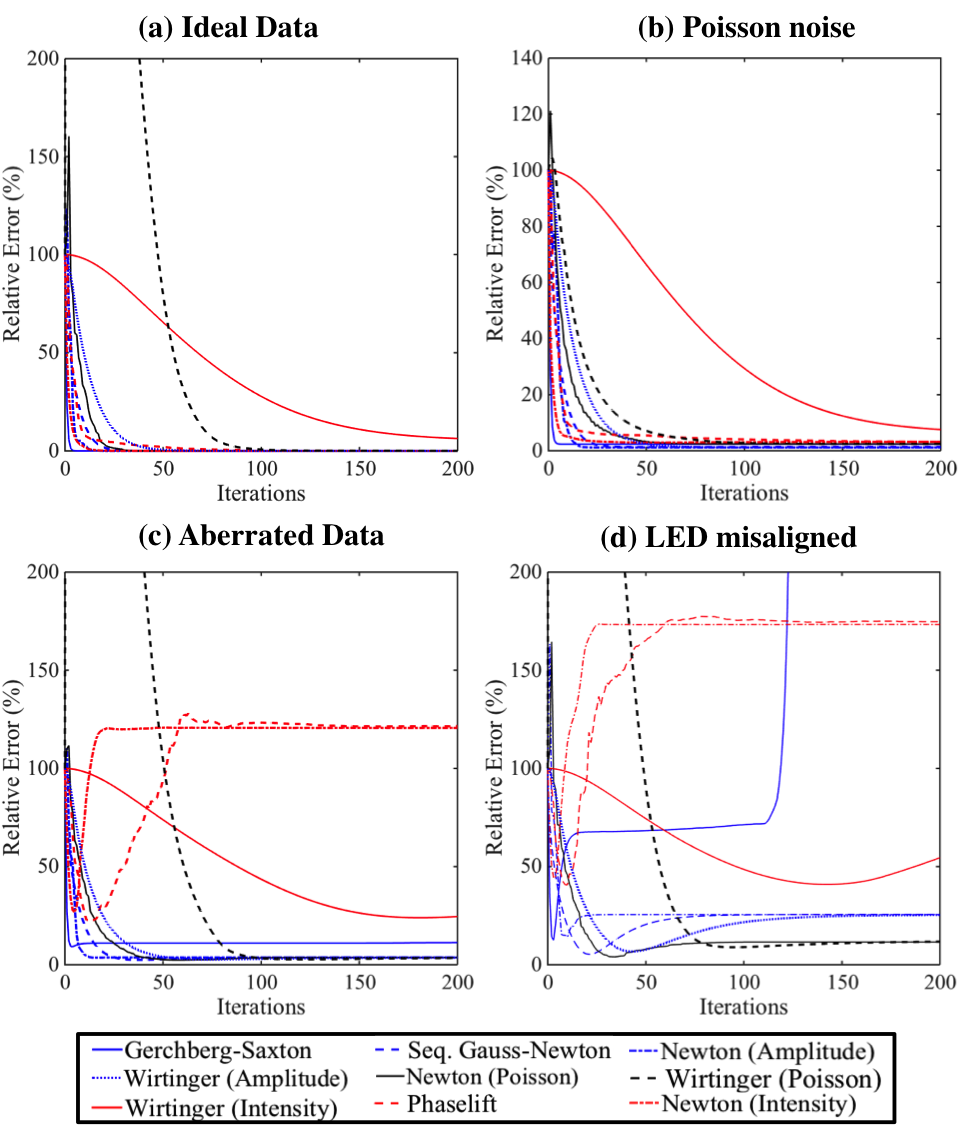}
\caption{Phase relative error as a function of iteration number for different algorithms with the (a) ideal data, (b) Poisson noise data, (c) aberrated data and (d) LED misaligned data. When the data is not perfect, some of the algorithms may not converge to a correct solution.}
\label{Fig_ex}
\end{figure}

To examine the convergence of each algorithm, Figure~\ref{Fig_ex} plots the error for each iteration when using the aberrated dataset and LED misaligned dataset with different algorithms. The \textit{intensity-based} algorithms (red curves) clearly do not converge to the correct solution and can incur large errors when the data is not perfect. Compared to PhaseLift and the intensity-based Newton's method, the Wirtinger-flow algorithm seems to have lower error; however, this is only due to its slow divergence. If run for many iterations, it will eventually settle on a similarly error-corrupted result as the other two \textit{intensity-based} algorithms (not shown). We also observe that \emph{amplitude-based} (blue curves) and \textit{Poisson-likelihood-based} (black curve) algorithms converge to points with lower errors in a similar fashion. This behavior is well explained by the similarity of the algorithms in their use of gradients and Hessians (as shown in the Appendix). Again, the exception to the trend is the first-order Gerchberg-Saxton algorithm, which recovers the object fairly well with aberrated data, but goes unstable in the case of LED misalignment. Note that, when there is no pupil estimation step, the only difference between the Gerchberg-Saxton and the sequential Gauss-Newton algorithm is the step size. Since the latter algorithm gives a good reconstruction, while the former diverges, we conclude that the Gerchberg-Saxton step size is too large for a stable update in this particular case. 

\begin{table}[h]
\vspace{-2mm}
\centering
\caption{Convergence Speed}
\begin{tabular}{|c|c|c|c|c|}
\hline
\multirow{2}{*}{} & \multicolumn{2}{|c|}{Ideal data} & \multicolumn{2}{|c|}{Misaligned data}\\ \cline{2-5}
\multirow{2}{*}{} & \makecell{Iteration \\ number}  & Runtime (s) & \makecell{Iteration \\ number}  & Runtime (s) \\ \hline
\makecell{Gerchberg \\ Saxton} & 4 & 2.22  & diverges & diverges \\ \hline
\makecell{Sequential \\ Gauss-Newton} & 23 & 12.97 & 83 & 46.8 \\ \hline
\makecell{Amplitude \\ Newton} & 13 & 100.49 & 20 & 154.6 \\ \hline
\makecell{Amplitude \\ Wirtinger} & 46 & 26.28 & 158 & 89.52 \\ \hline
\makecell{Poisson \\ Newton} & 28 & 211.68 & 77 & 582.1 \\ \hline 
\makecell{Poisson \\ Wirtinger} & 96 & 54.46 & 153  & 87.36 \\ \hline 
\makecell{Intensity \\ Wirtinger} & 1481 & 651.64 & diverges & diverges \\ \hline 
\makecell{PhaseLift}  & 67 & 386.28 & diverges & diverges \\ \hline
\makecell{Intensity \\ Newton} & 12 & 74.44 & diverges & diverges \\ \hline 
\end{tabular}
\label{tab_speed}
\vspace{-3mm}
\end{table}

The convergence speed of each algorithm can be determined from Figure~\ref{Fig_ex} using two metrics: \textit{number of iterations} required and total \textit{runtime}. We choose the convergence curves from the cases of ideal data and LED misaligned data and compare their iteration numbers and runtimes in Table~\ref{tab_speed}. All the algorithms were implemented in MATLAB on an Intel i7 2.8 GHz CPU computer with 16G DDR3 RAM under OS X operating system. We define convergence as the point when the relative phase error reaches its stable point. The comparison does not consider the divergent cases. In the ideal data case, we can see that the sequential methods outperform all the other algorithms in terms of runtime. The Gerchberg-Saxton algorithm is the fastest in terms of both iteration number and runtime for this perfect dataset. The global Newton's method using \textit{intensity-based} and \textit{amplitude-based} cost functions also converge very fast in terms of iteration number. The Wirtinger flow algorithm takes much longer to reach convergence both in runtime and iteration number. For the case of the LED misaligned data, only five algorithms converge. In terms of iteration number, the \textit{amplitude-based} Newton's method converges much faster than the other four, as expected. However, the sequential Gauss-Newton algorithm converges much faster in terms of the runtime. Though the global Newton's method is theoretically better than the others, it takes significant time to calculate the full Hessian matrix. Thus, the sequential Gauss-Newton method is our preferred algorithm in practice, because it provides excellent robustness while also enabling fast runtimes and reasonable computational complexity.

The main conclusions to be drawn from this section are that the FPM optimization algorithms which are formulated from \textit{amplitude-based} and \textit{Poisson-likelihood-based }cost functions are more tolerant to imperfect datasets with both Poisson noise and physical deviations like model mis-match, which were represented by aberrations and LED misalignment here. In the next section, we will explain more about the causes for this trend.

\subsection{Noise Model Analysis}

\begin{figure}[H]
\centering
\vspace{-3mm}
\includegraphics[width=13cm]{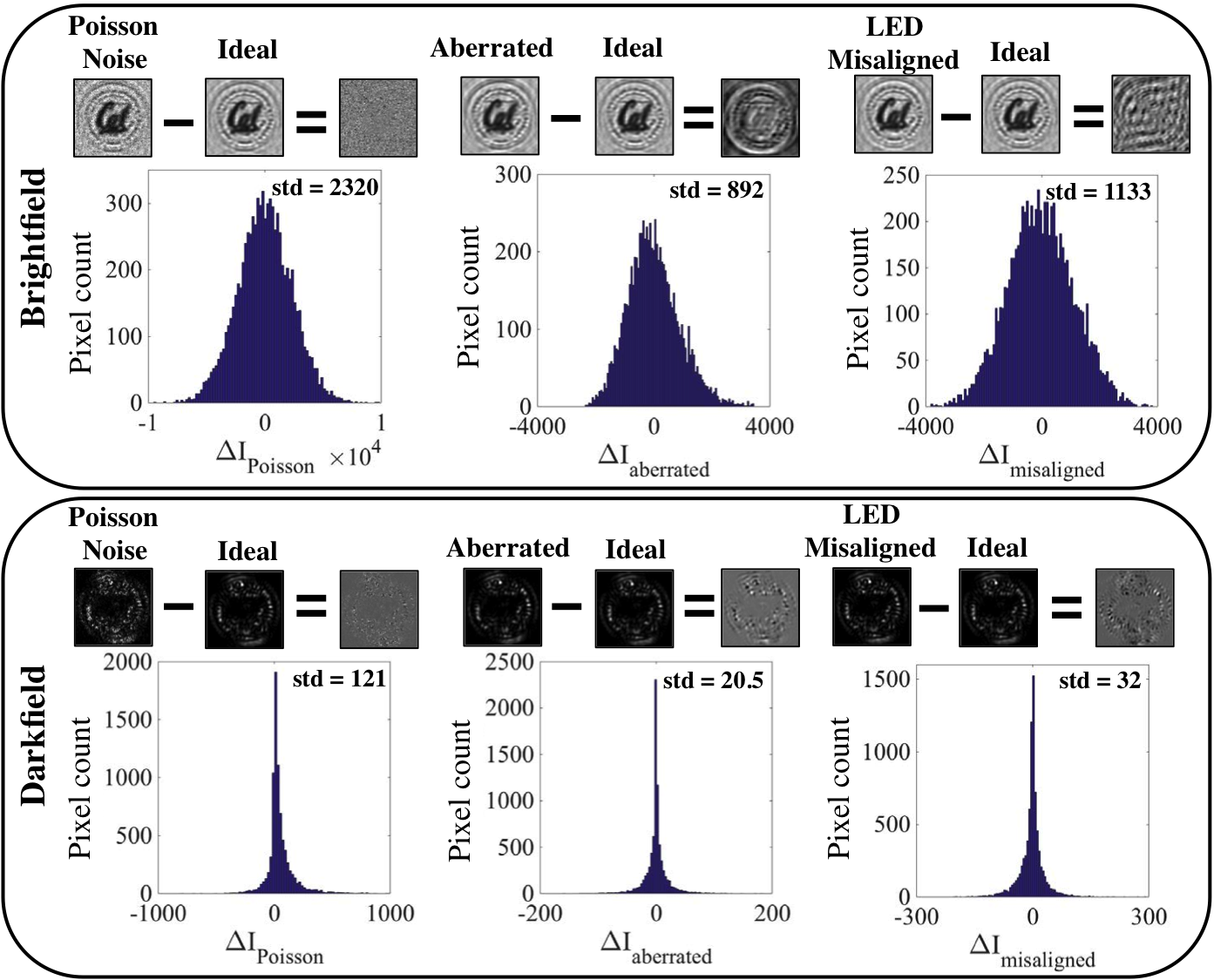}
\vspace{-2mm}
\caption{Both Poisson noise and model mis-match (aberrations, LED misalignment) cause errors that scale with mean intensity. Here, histograms show the intensity deviations under Poisson noise, aberration, and misalignment for a brightfield and darkfield image.}
\label{Fig6_2}
\end{figure}

The reason why \textit{amplitude-based} and \textit{Poisson-likelihood-based} algorithms have superior tolerance to experimental errors is due to their Poisson noise model. Each of these algorithms makes an implicit or explicit assumption that the magnitude of the errors in the data scale with the measured intensity. This is obviously a good model for Poisson noise errors, which are defined as noise which scales with intensity. It is not as obvious that the model mis-match errors (aberrations and LED misalignment) scale with intensity as well. To demonstrate this, Fig.~\ref{Fig6_2} shows the histogram of the difference between the deviated dataset and the ideal dataset, for the cases of both brightfield and darkfield images. The histograms show a similar trend - all of the brightfield errors are much larger than the darkfield errors, with a similar statistical variation. Thus, the errors from Poisson noise, aberrations \textit{and} LED misalignment all scale with the measured intensity. In our experimental data, there are always aberrations in the objective lens, LED misalignment, and Poisson shot noise. Since the noise model for the \textit{amplitude-based} and \textit{Poisson-likelihood-based} algorithms match the actual noise properties, these algorithms perform better than the \textit{intensity-based} algorithms. And since the images captured by FPM have drastically different intensity values, this effect dominates the reconstruction artifacts. Note that these large variations in intensity values are specific to FPM and likely do not play a major role in other phase imaging schemes (e.g. phase from defocus or traditional ptychography), where images do not have such a wide range of intensity values. In our experiments, the Poisson noise is fairly low (due to use of a high-performance sCMOS sensor), but the model mismatch in the experimental data can cause effects similar to strong Poisson noise.

\begin{figure}[H]
\centering
\vspace{-3mm}
\includegraphics[width=14cm]{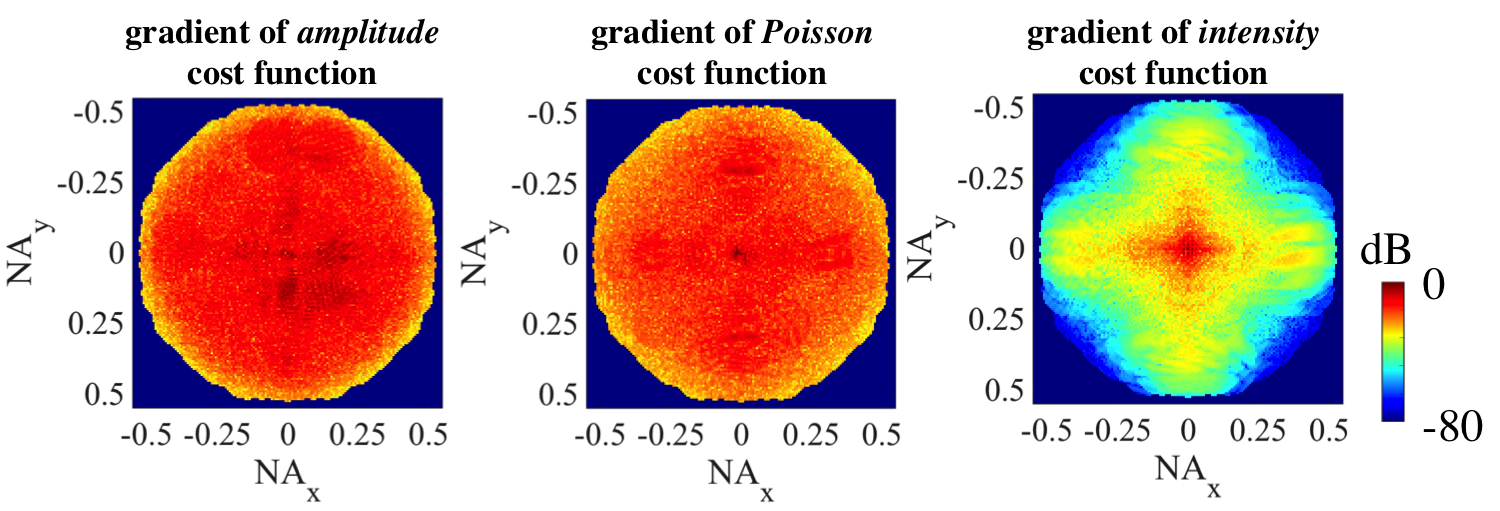}
\vspace{-2mm}
\caption{The \textit{intensity-based} cost function gives higher weighting to images in the low spatial frequency region of the Fourier domain, resulting in high-frequency artifacts. Here, we show the gradient of the \textit{amplitude-based}, \textit{Poisson-likelihood-based} and \textit{intensity-based} cost functions at the tenth iteration, using experimental data.}
\label{Fig7_2}
\end{figure}

For further understanding, we look closer at the relationship between the noise model and the cost function. Our optimization algorithms are derived from three cost functions. Each of the cost functions makes a noise model assumption. The \textit{intensity-based} cost function assumes that noise in the data follows a white Gaussian noise model, which means that the standard deviation of the noise is assumed to be the same across the brightfield and darkfield images. Recall that the standard deviation of a Gaussian noise probability model is related to the weight in the cost function for each pixel, as shown in Eq.~\ref{eqn_Gaussian_log}. The larger the standard deviation (amount of noise) at any pixel in Fourier space, the smaller the weighting, since noisy pixels should be trusted less. In the Gaussian noise model, the weights in the cost function for large-valued pixels and small-value pixels are the same. However, the deviation for brightfield images is much larger than that for darkfield images, as shown in Fig.~\ref{Fig6_2}. Hence, the brightfield images will contribute more to the total cost function value if the weights are all the same, due to their high intensity. The result is that the \textit{intensity-based} (Gaussian noise model) algorithms focus mostly on the brightfield images, which correspond to low spatial frequency information, and the darkfield images do not contribute much. The result is a failure in the high-frequency reconstruction, as we saw in Figs.~\ref{Fig4},~\ref{Fig2},~\ref{Fig3}, and loss of effective resolution since the darkfield images contain all the sub-diffraction-limit information. To illustrate the dramatic difference in weights, Fig.~\ref{Fig7_2} shows the gradient of the different cost functions. Obviously, the intensity cost function gives much higher weighting to low spatial frequencies, which causes the high-frequency artifacts. 

Since the amplitude-based cost function shares a similar gradient and Hessian with the Poisson likelihood function, as shown in the Appendix and Fig.~\ref{Fig7_2}, it is not surprising that they both produce a similar quality reconstruction. Both of these cost functions assume the noise in the data follow a Poisson distribution, with the standard deviation scaling with the measured intensity. This assumption matches the actual error better than the white Gaussian assumption. The actual noise or deviations in the experiments for brightfield images have larger standard deviation, while that for darkfield images have smaller standard deviation. Under the Poisson noise model, the weight in the cost function is smaller for the noisy brightfield images and larger for the darkfield images. At the end, algorithms based on the Poisson noise model put more emphasis on the darkfield images and thus get a better reconstruction compared to the \textit{intensity-based} algorithms. Figure~\ref{Fig7_2} shows that the gradients for the \textit{amplitude-based} and \textit{Poisson-likelihood-based} cost function are similar and are more uniform throughout the whole Fourier space.

\subsection{Pupil estimation} \label{sec: Pupil}

There are already more sophisticated FPM extensions to correct for some model mis-match errors~\cite{Ou:14, Tian2014}, similar to the probe correction algorithms in traditional ptychography~\cite{thibault2009probe}. Both of the methods previously developed for Fourier ptychography are derived from the amplitude-based formulation. By taking the derivative of the cost function with respect to $\mathbf{P}$, the decent direction to estimate the pupil function can be calculated as
\begin{equation}
\nabla_\mathbf{P} f_{A,\ell}(\mathbf{O},\mathbf{P}) = -{\rm diag}(\bar{\mathbf{Q}}_\ell \bar{\mathbf{O}}) \left[ \mathbf{F} {\rm diag} \left( \frac{\sqrt{\mathbf{I}_\ell}}{|\mathbf{g}_\ell|}\right) \mathbf{g}_\ell - {\rm diag} (\mathbf{P}) \mathbf{Q}_\ell \mathbf{O} \right].
\label{eqn_grad_Pupil}
\end{equation}

By applying the pupil estimation step after each object estimation using this gradient or approximated Hessian, the sequential gradient descent~\cite{Ou:14} and the sequential Gauss-Newton method~\cite{Tian2014} including pupil estimation can be derived. Here we only consider the amplitude-based cost function, for simplicity. 

\begin{figure}[H]
\centering
\vspace{-3mm}
\includegraphics[width=10cm]{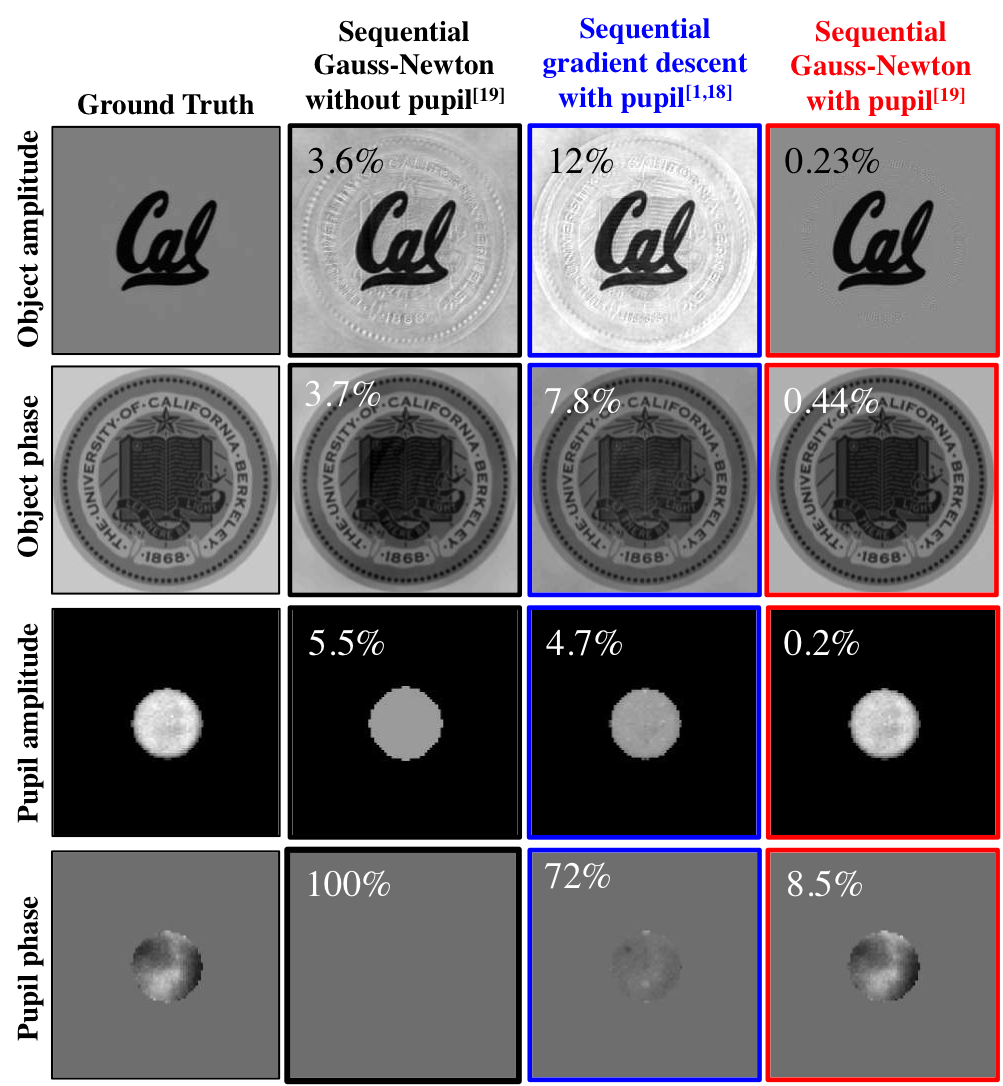}
\vspace{-2mm}
\caption{Object and pupil reconstruction results using different algorithms, with and without pupil estimation. The second-order method (sequential Gauss-Newton) with pupil estimation gives the best result, as expected. In this case, we find that the second-order method \textit{without} pupil estimation is already better than first-order method (sequential gradient descent) \textit{with} pupil estimation.}
\label{Fig9}
\vspace{-3mm}
\end{figure}

We wish to investigate the improvements obtained by adding a pupil estimation step to both first and second-order optimization algorithms. Figure~\ref{Fig9} shows the reconstruction result from the sequential gradient descent (first-order) and sequential Gauss-Newton (second-order) algorithms, using the aberrated dataset from the previous simulations. The numbers at the top left corner are the relative error compared to the ground truth simulated image. As can be seen, adding the pupil estimation step gives a better complex-field reconstruction, and the second-order (Gauss-Newton) method with pupil estimation provides the best result. 

Surprisingly, however, the second-order reconstruction without pupil estimation is better than the first-order reconstruction with pupil estimation, for this case. This highlights the robustness to aberrations that a second-order optimization scheme enables. The second-order nature of the algorithm makes it faster in convergence, and also more stable. In terms of runtime, the pupil estimation step takes about the same time as the object reconstruction part, so the algorithm is two times slower when the pupil function step is incorporated. 

\subsection{Misalignment correction}

Another possible correction scheme for model mis-match is that for LED misalignment. Since each LED position corresponds to a certain shift of the pupil function in the Fourier domain, this is similar to the shift of the probe function in traditional ptychography. There, iterative algorithms have been proposed to correct for the positioning error of the probe function~\cite{guizar2008phase, Maiden2012annealing, Fucai:2013, Ashish:2014OE}. In~\cite{guizar2008phase,Ashish:2014OE}, a gradient of the cost function with respect to the shift of the probe function has been calculated and the conjugate gradient method has been applied to correct for the positioning error. In~\cite{Maiden2012annealing}, a simulated annealing method is adopted to estimate the shift of the probe function. The simulated annealing method is also adopted to correct for the misalignment of the spatial light modulator in a overlapped Fourier coding system~\cite{Horstmeyer2014}, analogous to FPM. In our experiments, we observe that the simulated annealing method can locate the LED positions more accurately than other methods. Thus, we only compare with the simulated annealing method.

Simulated annealing is a method of searching unknown variables over a finite space to minimize or maximize the function of merit - the cost function in our case. Instead of exhaustively testing all the possible states, simulated annealing iteratively approaches the optimal state. At the first iteration, the algorithm randomly searches several states in the space and selects the one with the smallest cost function value. The algorithm then starts at this state for the next iteration, slowly reducing the search range in the following iterations until convergence. 

\begin{figure}[H]
\centering
\includegraphics[width=14.2cm]{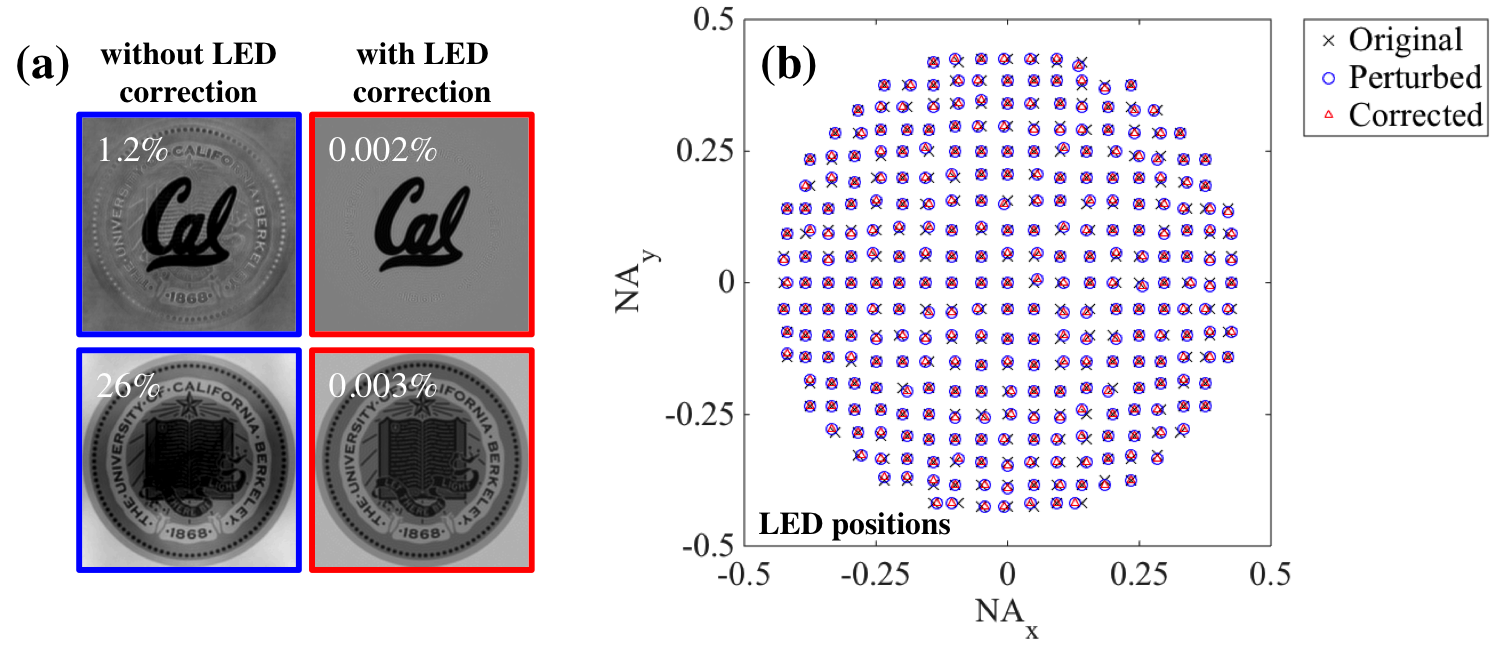}
\vspace{-4mm}
\caption{ (a) Adding LED misalignment correction improves the reconstruction results (sequential Gauss-Newton method). (b) The original, perturbed, and corrected LED positions in angular coordinates. LED correction accurately retrieves the actual LED positions.}
\label{Fig11}
\vspace{-3mm}
\end{figure}

In our sequential algorithm, the whole optimization problem is divided into many sub-optimization problems for different collected images. At each sub-optimization problem, a gradient descent or Gauss-Newton method is applied to update that corresponding region in Fourier domain. To add a LED mis-alignment correction step, the simulated annealing algorithm can be incorporated into each sub-iteration to find an optimal shift of the pupil function. In each sub-iteration, the down-sampling matrix, $\mathbf{Q}_\ell$, which contains the information of the pupil shift, is tested according to the annealing process for several possible states corresponding to different shifts of the pupil. The state with the smallest cost function value is selected to update the old down-sampling matrix. Then, the new down-sampling matrix is used to update the corresponding region in the Fourier domain. 

The simulated annealing method estimates the LED positions with good accuracy. Figure~\ref{Fig11} shows the reconstruction result from the simulated LED misaligned dataset, both with and without the LED correction step. The result using the LED correction clearly shows better quality and smaller error, as seen in Fig.~\ref{Fig11}(a). Since the LED correction scheme also estimates the actual LED positions, which we intentionally perturbed in order to impose a known error, we can also compare the actual and recovered LED positions, shown in Fig.~\ref{Fig11}(b).  

\begin{figure}[H]
\centering
\includegraphics[width=14.5cm]{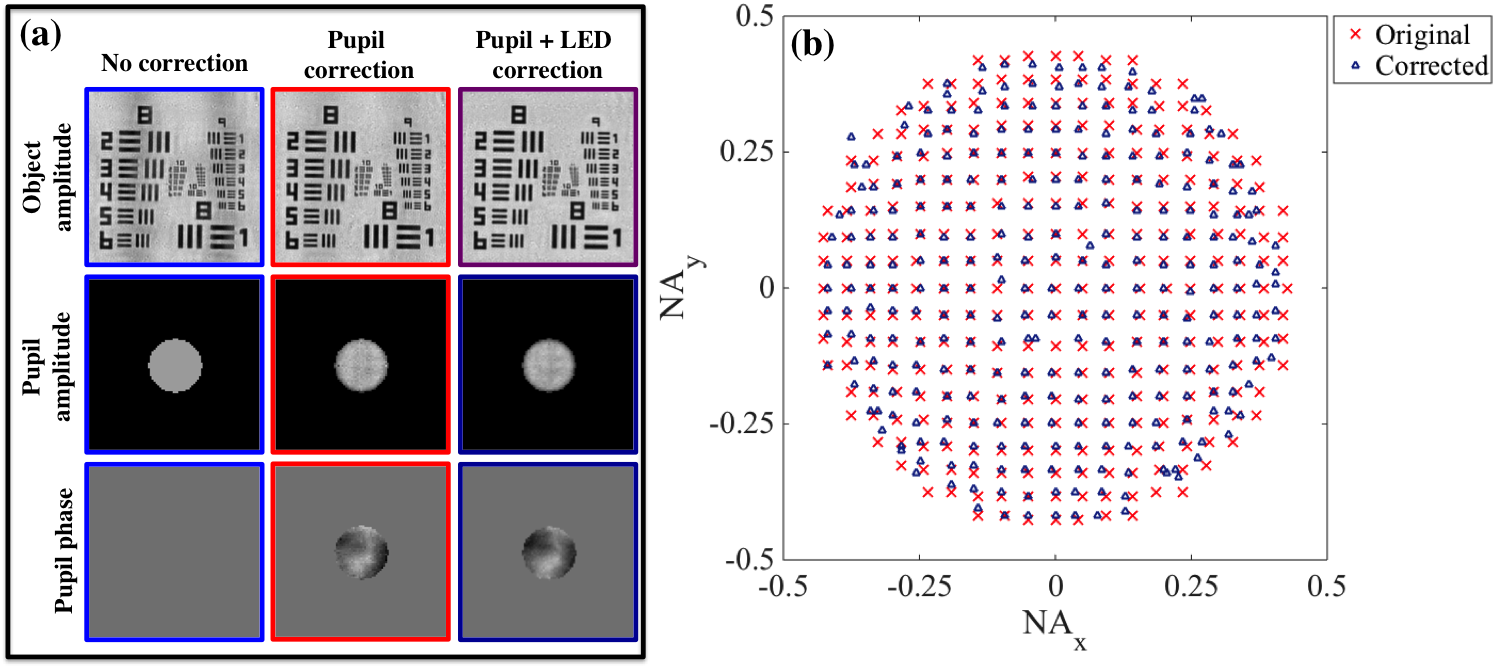}
\vspace{-5mm}
\caption{Experimental reconstructions with and without LED misalignment correction (sequential Gauss-Newton method). (a) The reconstructed object and pupil. (b) The original and corrected LED positions, in angular coordinates.}
\label{Fig12}
\vspace{-3mm}
\end{figure}

To complete the picture, we now show experimental reconstructions with and without the two correction schemes: pupil correction and LED mis-alignment corrections (see Fig.~\ref{Fig12}). Since we do not know ground truth for our experiments, we can only make qualitative observations. An incremental improvement is observed when adding the pupil estimation and then the LED correction steps - the background variation becomes flatter. Figure~\ref{Fig12}(b) shows the corrected LED positions compared to the original ones, in angular coordinates. Corrected positions of LEDs in different regions share similar offset because the fabrication process of the LED array can cause unexpected position misalignment for each LED. Notice that the LEDs at the edges (corresponding to higher angles of illumination) incur more variation, since these are more sensitive to calibration. Also, many of the large deviations occur at the edges that are not along the horizontal and vertical axes. In these areas, the LED position recovery is poor because the object has very little information there (the resolution test target contains only square features) and so the data contains little information about these areas. However, any errors in LED positions in this area will also not significantly affect the reconstruction if they do not contribute much energy to the object spectrum. If the goal was not to correct the image results, but rather to find the LED positions accurately, then one should choose an object that contains uniformly distributed spatial frequencies (e.g. a random diffuser or speckle field). Although the simulated annealing further improves our reconstruction, we note that it is more than ten times slower to process the data because of the local search performed at each sub-iteration. 

\section{Conclusion}

We formulated the Fourier ptychographic phase retrieval problem using maximum likelihood optimization theory. Under this framework, we reviewed the existing FPM algorithms and classified them based on their cost functions: \textit{amplitude-based} algorithms (akin to a Poisson noise model) and \textit{intensity-based} algorithms (akin to a white Gaussian noise model). We also derived a new algorithm based on the Poisson likelihood function, which is better suited for dealing with measurement imperfections. We compared the tolerance of these algorithms under errors due to experimental noise and model mis-match (aberrations and LED mis-alignment) using both simulated data and experimental data.  Because the noise and model mis-match error for brightfield and darkfield images depend on the measured intensity, the \textit{amplitude-based} and \textit{Poisson-likelihood-based} algorithms from the Poisson noise model are more robust than the \textit{intensity-based} algorithms. This can be explained by the standard deviation of the noise model determining the weight of each image in the optimization. Hence, \textit{intensity-based} algorithms over-weight the brightfield images, resulting in poor high-frequency reconstruction.  

We used existing pupil estimation algorithms and proposed a simulated-annealing-based LED correction algorithm to algorithmically fix the experimental deviations. We compared the performance of the pupil estimation algorithms and found that second-order methods give the best results. We also showed the capability of the simulated annealing method to correct for misaligned LEDs and find their actual positions. 

Based on our studies, we conclude that the global Newton's method gives the best reconstruction, but may have high computational cost. Considering both robustness and computational efficiency, we find that sequential Gauss-Newton method provides the best trade-offs for large-scale applications. Its experimental robustness is verified in our recent time-series \emph{in vitro} experiments~\cite{Tian2015a}. Our open source code for this algorithm can be downloaded at \cite{opensource}.

\section*{Acknowledgments}
Funding was provided by the Gordon and Betty Moore Foundation's Data-Driven Discovery Initiative through Grant GBMF4562 to Laura Waller (UC Berkeley).

\section*{Appendix A: Gradient and Hessian calculation}
\subsection*{Gradient:}
Consider that equations (\ref{eqn_amplitude_rew}) and (\ref{eqn_intensity_rew}) can be expressed as 
\begin{eqnarray}
&&f_A(\mathbf{O}) = \sum_\ell \mathbf{f}_{A\ell}^\dagger \mathbf{f}_{A\ell} \nonumber \\
&&f_I(\mathbf{O}) = \sum_\ell \mathbf{f}_{I\ell}^\dagger \mathbf{f}_{I\ell},
\label{eqn_ff}
\end{eqnarray}
where $\mathbf{f}_{A\ell} \equiv \sqrt{\mathbf{I}_\ell} - |\mathbf{g}_\ell|$, and $\mathbf{f}_{I\ell} \equiv \mathbf{I}_\ell - |\mathbf{g}_\ell|^2$. 

Then, calculate the derivative of $f_A$ with respect to $\mathbf{O}$, and it can then be expressed as
\begin{eqnarray}
&&\nabla_{\mathbf{O}} f_A(\mathbf{O})  = \sum_\ell \left [\frac{\partial(\mathbf{f}_{A\ell}^\dagger \mathbf{f}_{A\ell})}{\partial\mathbf{O}}\right]^\dagger = \sum_\ell \left [\frac{\partial(\mathbf{f}_{A\ell}^\dagger \mathbf{f}_{A\ell})}{\partial\mathbf{f}_{A\ell}} \frac{\partial\mathbf{f}_{A\ell}}{\partial\mathbf{O}}\right]^\dagger.
\label{eqn_grad}
\end{eqnarray}
Using $|\mathbf{g}_\ell|^2 = {\rm diag}(\bar{\mathbf{g}}_\ell) \mathbf{g}_\ell$ and $|\mathbf{g}_\ell| = (|\mathbf{g}_\ell|^2)^{1/2}$, two chain rule parts in (\ref{eqn_grad}) are calculated as 
\begin{eqnarray}
&&\frac{\partial(\mathbf{f}_{A\ell}^\dagger \mathbf{f}_{A\ell})}{\partial\mathbf{f}_{A\ell}} = 2 \mathbf{f}_{A\ell}^\dagger \nonumber \\
&&\frac{\partial \mathbf{f}_{A\ell}}{\partial \mathbf{O}} = -\frac{\partial (|\mathbf{g}_\ell|^2)^{1/2}}{\partial (|\mathbf{g}_\ell|^2)}\frac{\partial ({\rm diag}(\bar{\mathbf{g}}_\ell)\mathbf{g}_\ell)}{\partial \mathbf{O}} = -\frac{1}{2}{\rm diag}\left( \frac{\bar{\mathbf{g}}_\ell}{|\mathbf{g}_\ell|}\right) \mathbf{F}^{-1} {\rm diag}(\mathbf{P}) \mathbf{Q}_\ell,
\label{eqn_grad_two_part}
\end{eqnarray}
if $\mathbf{g}_\ell$ does not contain any zero entries for $\ell = 1, \ldots, N_{img}$. 

By plugging these two terms into (\ref{eqn_grad}), the gradient of $f_A$ with respect to $\mathbf{O}$ becomes 
\begin{eqnarray}
&&\nabla_{\mathbf{O}} f_A(\mathbf{O}) = - \sum_\ell \mathbf{Q}_\ell^\dagger  {\rm diag}(\bar{\mathbf{P}})  \mathbf{F} {\rm diag}\left(\frac{\mathbf{g}_\ell}{|\mathbf{g}_\ell|}\right)(\sqrt{\mathbf{I}_\ell} - |\mathbf{g}_\ell|) \nonumber \\
&&\hspace{0.5 in} = - \sum_\ell \mathbf{Q}_\ell^\dagger  {\rm diag}(\bar{\mathbf{P}}) \left(\mathbf{F} {\rm diag}\left(\frac{\sqrt{\mathbf{I}_\ell}}{|\mathbf{g}_\ell|}\right) \mathbf{g}_\ell - {\rm diag}(\mathbf{P})\mathbf{Q}_\ell\mathbf{O} \right).
\label{eqn_grad_fA}
\end{eqnarray}

The gradient for $f_I$ can be calculated in the similar way, and the chain rule part of $\mathbf{f}_{I\ell}$ can be calculated as
\begin{equation}
\frac{\partial\mathbf{f}_{I\ell}}{\partial\mathbf{O}} = -\frac{\partial ({\rm diag}(\bar{\mathbf{g}}_\ell) \mathbf{g}_\ell)}{\partial \mathbf{O}} = -{\rm diag}(\bar{\mathbf{g}}_\ell) \mathbf{F}^{-1} {\rm diag}(\mathbf{P}) \mathbf{Q}_\ell.
\label{eqn_grad_two_part_I}
\end{equation} 

With (\ref{eqn_grad_two_part_I}), it is clear to express the gradient of $f_I$ as 
\begin{equation}
\nabla_{\mathbf{O}} f_I(\mathbf{O}) = \sum_\ell \left [\frac{\partial(\mathbf{f}_{I\ell}^\dagger \mathbf{f}_{I\ell})}{\partial\mathbf{f}_{I\ell}} \frac{\partial\mathbf{f}_{I\ell}}{\partial\mathbf{O}}\right]^\dagger = -2 \sum_\ell \mathbf{Q}_\ell^\dagger  {\rm diag}(\bar{\mathbf{P}})  \mathbf{F} {\rm diag}(\mathbf{g}_\ell)(\mathbf{I}_\ell - |\mathbf{g}_\ell|^2).
\label{eqn_grad_fI}
\end{equation}

The calculation of gradient of $\mathcal{L}_{\rm Poisson} (\mathbf{O})$ with respect to $\mathbf{O}$ is different from the other two. With the expression (\ref{eqn_Poisson_rew}), the gradient of Poisson likelihood function can be calculated as
\begin{eqnarray}
&&\hspace{-0.9 in}\nabla_{\mathbf{O}} \mathcal{L}_{\rm Poisson} (\mathbf{O}) = \left(\frac{\partial \mathcal{L}_{\rm Poisson}}{\partial \mathbf{O}}\right)^\dagger  = \left( \sum_\ell \sum_j \left[ - \frac{I_{\ell,j}}{\mathbf{a}_{\ell,j}^\dagger \mathbf{O}}\mathbf{a}_{\ell,j}^\dagger + \mathbf{a}_{\ell,j}^\dagger  \right] \right)^\dagger \nonumber \\
&&= -\left(\sum_\ell \sum_j \left[I_{\ell,j} - \mathbf{a}_{\ell,j}^\dagger \mathbf{O} \right] \frac{1}{\mathbf{a}_{\ell,j}^\dagger \mathbf{O}} \mathbf{a}_{\ell,j}^\dagger \right)^\dagger \nonumber \\
&&= -\left( \sum_\ell (\mathbf{I}_\ell - |\mathbf{g}_\ell|^2)^\dagger {\rm diag}\left( \frac{1}{|\mathbf{g}_\ell|^2} \right) {\rm diag} (\bar{\mathbf{g}}_\ell) \mathbf{F}^{-1} {\rm diag} (\mathbf{P}) \mathbf{Q}_\ell \right)^\dagger \nonumber \\
&&= - \sum_\ell \mathbf{Q}_\ell^\dagger {\rm diag} (\bar{\mathbf{P}}) \mathbf{F} {\rm diag}\left( \frac{\mathbf{g}_\ell}{|\mathbf{g}_\ell|^2} \right) (\mathbf{I}_\ell - |\mathbf{g}_\ell|^2).
\label{eqn_grad_Poisson}
\end{eqnarray}
This is equivalent to the gradient of the \textit{intensity-based} cost function with added weight $1/|\mathbf{g}_\ell|^2$ to the component from each image. In addition, this gradient is very similar to that from the \textit{amplitude-based} cost function.

Since we have gradients for all cost functions, the updating equation for the gradient descent method can then be expressed as
\begin{equation}
\mathbf{O}^{(i+1)} = \mathbf{O}^{(i)} - \alpha^{(i)} \nabla_{\mathbf{O}} f(\mathbf{O}^{(i)}),
\label{eqn_grad_des}
\end{equation}
where $i$ denotes the iteration number, $\alpha$ is the step size chosen by the line search algorithm, and $f(\mathbf{O})$ can be either \textit{intensity-based} or \textit{amplitude-based} cost function. 

Looking at $\nabla_{\mathbf{O}} f_A(\mathbf{O})$, $\nabla_{\mathbf{O}} f_I(\mathbf{O})$ and $\nabla_{\mathbf{O}} \mathcal{L}_{\rm Poisson}(\mathbf{O})$, they all contain the term $\mathbf{Q}_\ell^\dagger {\rm diag}(\bar{\mathbf{P}})$ following by a residual term. The residual term basically finds the difference between the estimation and the measurement. This difference carries the information to update the previous estimation. Since each measurement carries the information for a specific region in the Fourier space, the $\mathbf{Q}_\ell^\dagger {\rm diag}(\bar{\mathbf{P}})$ term brings this updating information back to the right place corresponding to some spatial frequency. For $\nabla_{\mathbf{O}} f_A(\mathbf{O})$, the first term in the residual shows the replacement of the amplitude in the real domain, which is the projection from the estimation to the modulus space. Thus, the gradient descent method using the \textit{amplitude-based} cost function is similar to the projection-based phase retrieval solver. 

\subsection*{Hessian:}

The second-order Taylor expansion on an arbitrary real function $f(\mathbf{c})$ with a complex vector $\mathbf{c} = (\mathbf{O}^T,\bar{\mathbf{O}}^T)^T$ at certain point $\mathbf{c}_0 = (\mathbf{O}_0^T,\bar{\mathbf{O}}_0^T)^T$ can be written as~\cite{Ken2009}
\begin{equation}
f(\mathbf{c}) \approx f(\mathbf{c}_0) +  \nabla f(\mathbf{c}_0)^\dagger (\mathbf{c} - \mathbf{c}_0) + \frac{1}{2}(\mathbf{c} - \mathbf{c}_0)^\dagger \mathbf{H}_{\mathbf{c}\mathbf{c}}(\mathbf{c}_0) (\mathbf{c} - \mathbf{c}_0),
\label{eqn_taylor}
\end{equation}
where the matrix $\mathbf{H}_{\mathbf{c}\mathbf{c}}$ is the Hessian of $f(\mathbf{c})$. For the case of a single-value function, the second-order term in the Taylor expansion denotes the curvature of the function at that expansion point. Thus, this Hessian matrix similarly contains the curvature information of the original multi-variate function. 

If the Hessian is a diagonal matrix, each diagonal entry denotes the curvature in each corresponding dimension. If the Hessian is not diagonal, a coordinate transformation can be found to make the Hessian diagonal by using eigenvalue decomposition. For a convex problem, the Hessian is positive semidefinite. The curvatures of the cost function in different dimensions are always nonnegative. A standard optimization process can lead to a global minimum. However, if the problem is non-convex, a standard optimization process will probably lead to a local minimum. Calculating the Hessian of a cost function is useful either to examine the optimization process or to speed up the convergence rate by using Newton's method. 

From~\cite{Yang:2011ty,Ken2009}, the definition for the Hessian of a real-value function with multiple complex variables is a $2n^2 \times 2n^2$ matrix and  can be expressed as 
\begin{eqnarray}
&&\mathbf{H}_{\mathbf{c}\mathbf{c}} = 
\begin{bmatrix}
\mathbf{H}_{\mathbf{O}\mathbf{O}} & \mathbf{H}_{\bar{\mathbf{O}}\mathbf{O}} \\
\mathbf{H}_{\mathbf{O}\bar{\mathbf{O}}} & \mathbf{H}_{\bar{\mathbf{O}}\bar{\mathbf{O}}} 
\end{bmatrix},
\label{eqn_def_hessian}
\end{eqnarray}
where each component $n^2\times n^2$ matrices can be further calculated as 
\begin{eqnarray}
&&\mathbf{H}_{\mathbf{O}\mathbf{O}} = \frac{\partial}{\partial\mathbf{O}}\left( \frac{\partial f}{\partial\mathbf{O}} \right)^\dagger, \mathbf{H}_{\bar{\mathbf{O}}\mathbf{O}} = \frac{\partial}{\partial\bar{\mathbf{O}}}\left( \frac{\partial f}{\partial\mathbf{O}} \right)^\dagger \nonumber \\
&&\mathbf{H}_{\mathbf{O}\bar{\mathbf{O}}} = \frac{\partial}{\partial\mathbf{O}}\left( \frac{\partial f}{\partial\bar{\mathbf{O}}} \right)^\dagger, \mathbf{H}_{\bar{\mathbf{O}}\bar{\mathbf{O}}} = \frac{\partial}{\partial\bar{\mathbf{O}}}\left( \frac{\partial f}{\partial\bar{\mathbf{O}}} \right)^\dagger.
\label{eqn_def_hessian2}
\end{eqnarray}
Similar to the calculation of the gradient, the components of the Hessians for \textit{amplitude-based}, \textit{intensity-based}, and \textit{Poisson-likelihood-based} cost functions can be calculated by taking an additional derivative on the gradient of the cost functions. The components of the Hessian for the amplitude-based cost function are
\begin{eqnarray}
&&\mathbf{H}^A_{\mathbf{O}\mathbf{O}} = \sum_\ell\mathbf{Q}_\ell^\dagger {\rm diag}(\bar{\mathbf{P}})\mathbf{F} \left[\mathbf{1}-\frac{1}{2}{\rm diag}\left(\frac{\sqrt{\mathbf{I}_\ell}}{|\mathbf{g}_\ell|}\right)\right] \mathbf{F}^{-1} {\rm diag}(\mathbf{P})\mathbf{Q}_\ell \nonumber \\
&&\mathbf{H}^A_{\bar{\mathbf{O}}\mathbf{O}} = \frac{1}{2}\sum_\ell \mathbf{Q}_\ell^\dagger {\rm diag}(\bar{\mathbf{P}})\mathbf{F} {\rm diag}\left(\frac{\sqrt{\mathbf{I}_\ell}\mathbf{g}_\ell^2}{|\mathbf{g}_\ell|^3}\right) \bar{\mathbf{F}}^{-1} {\rm diag}(\bar{\mathbf{P}})\bar{\mathbf{Q}}_\ell \nonumber \\
&&\mathbf{H}^A_{\mathbf{O}\bar{\mathbf{O}}} = \frac{1}{2}\sum_\ell \mathbf{Q}_\ell^T {\rm diag}(\mathbf{P})\bar{\mathbf{F}} {\rm diag}\left(\frac{\sqrt{\mathbf{I}_\ell}\bar{\mathbf{g}}_\ell^2}{|\mathbf{g}_\ell|^3}\right) \mathbf{F}^{-1} {\rm diag}(\mathbf{P})\mathbf{Q}_\ell \nonumber \\
&&\mathbf{H}^A_{\bar{\mathbf{O}}\bar{\mathbf{O}}} = \sum_\ell \mathbf{Q}_\ell^T {\rm diag}(\mathbf{P})\bar{\mathbf{F}} \left[\mathbf{1}-\frac{1}{2}{\rm diag}\left(\frac{\sqrt{\mathbf{I}_\ell}}{|\mathbf{g}_\ell|}\right)\right] \bar{\mathbf{F}}^{-1} {\rm diag}(\bar{\mathbf{P}})\bar{\mathbf{Q}}_\ell,
\label{eqn_hessian_A}
\end{eqnarray}
where $\mathbf{1}$ is the $m^2\times m^2$ identity matrix. 

Likewise, the Hessian of the intensity-based cost function is 
\begin{eqnarray}
&&\mathbf{H}^I_{\mathbf{O}\mathbf{O}} = 2\sum_\ell\mathbf{Q}_\ell^\dagger {\rm diag}(\bar{\mathbf{P}})\mathbf{F} {\rm diag}(2|\mathbf{g}_\ell|^2 - \mathbf{I}_\ell) \mathbf{F}^{-1} {\rm diag}(\mathbf{P})\mathbf{Q}_\ell \nonumber \\
&&\mathbf{H}^I_{\bar{\mathbf{O}}\mathbf{O}} = 2\sum_\ell \mathbf{Q}_\ell^\dagger {\rm diag}(\bar{\mathbf{P}})\mathbf{F} {\rm diag}(\mathbf{g}_\ell^2) \bar{\mathbf{F}}^{-1} {\rm diag}(\bar{\mathbf{P}})\bar{\mathbf{Q}}_\ell \nonumber \\
&&\mathbf{H}^I_{\mathbf{O}\bar{\mathbf{O}}} = 2\sum_\ell \mathbf{Q}_\ell^T {\rm diag}(\mathbf{P})\bar{\mathbf{F}} {\rm diag}(\bar{\mathbf{g}}_\ell^2) \mathbf{F}^{-1} {\rm diag}(\mathbf{P})\mathbf{Q}_\ell \nonumber \\
&&\mathbf{H}^I_{\bar{\mathbf{O}}\bar{\mathbf{O}}} = 2\sum_\ell \mathbf{Q}_\ell^T {\rm diag}(\mathbf{P})\bar{\mathbf{F}} {\rm diag}(2|\mathbf{g}_\ell|^2 - \mathbf{I}_\ell) \bar{\mathbf{F}}^{-1} {\rm diag}(\bar{\mathbf{P}})\bar{\mathbf{Q}}_\ell.
\label{eqn_hessian_I}
\end{eqnarray}

Finally, the Hessian of the Poisson likelihood cost function is
\begin{eqnarray}
&&\mathbf{H}^P_{\mathbf{O}\mathbf{O}} = \sum_\ell \mathbf{Q}_\ell^\dagger {\rm diag} (|\mathbf{P}|^2) \mathbf{Q}_\ell \nonumber \\
&&\mathbf{H}^P_{\bar{\mathbf{O}}\mathbf{O}} = \sum_\ell \mathbf{Q}_\ell^\dagger {\rm diag} (\bar{\mathbf{P}}) \mathbf{F} {\rm diag}\left( \frac{\mathbf{I}_\ell \mathbf{g}_\ell^2}{|\mathbf{g}_\ell|^4}\right) \bar{\mathbf{F}}^{-1} {\rm diag} (\bar{\mathbf{P}}) \bar{\mathbf{Q}}_\ell \nonumber \\
&&\mathbf{H}^P_{\mathbf{O}\bar{\mathbf{O}}} = \sum_\ell \mathbf{Q}_\ell^T {\rm diag} (\mathbf{P}) \bar{\mathbf{F}} {\rm diag}\left( \frac{\mathbf{I}_\ell \bar{\mathbf{g}}_\ell^2}{|\mathbf{g}_\ell|^4}\right) \mathbf{F}^{-1} {\rm diag} (\mathbf{P}) \mathbf{Q}_\ell \nonumber \\
&&\mathbf{H}^P_{\bar{\mathbf{O}}\bar{\mathbf{O}}} = \sum_\ell \mathbf{Q}_\ell^T {\rm diag} (|\mathbf{P}|^2) \bar{\mathbf{Q}}_\ell.
\label{eqn+hessian_P}
\end{eqnarray}

In general, Newton's method, which is the second-order method using the inversion of Hessian matrix, is preferred in solving nonlinear least square problems because of its fast convergence and stability compared to the first-order methods such as gradient descent. The updating equation for Newton's method can be expressed as
\begin{eqnarray}
\begin{bmatrix}
\mathbf{O}^{(i+1)} \\
\bar{\mathbf{O}}^{(i+1)}
\end{bmatrix} = 
\begin{bmatrix}
\mathbf{O}^{(i)} \\
\bar{\mathbf{O}}^{(i)}
\end{bmatrix} - \alpha^{(i)} \mathbf{H}_{\mathbf{c}\mathbf{c}}^{-1} 
\begin{bmatrix}
\nabla_{\mathbf{O}}f(\mathbf{O}^{(i)}) \\
\nabla_{\bar{\mathbf{O}}}f(\mathbf{O}^{(i)})
\end{bmatrix}.
\label{eqn_Newton}
\end{eqnarray}


\end{document}